# Hazards Induced by Breach of Liquid Rocket Fuel Tanks: Conditions and Risks of Cryogenic Liquid Hydrogen-Oxygen Mixture Explosions

Viatcheslav (Slava) Osipov[1], Cyrill Muratov[2], Halyna Hafiychuk[3],
Ekaterina Ponizovskaya-Devine[4], Vadim Smelyanskiy[5]

*Applied Physics Group, Intelligent Systems Division, NASA Ames Research Center, Moffett Field, CA, 94035*

Donovan Mathias[6], Scott Lawrence[7], and Mary Werkheiser[8]
*Supercomputing Division, NASA Ames Research Center*, MS 258-1, Moffett Field, CA, 94035, USA

**We analyze the data of purposeful rupture experiments with LOx and LH2 tanks, the Hydrogen-Oxygen Vertical Impact (HOVI) tests that were performed to clarify the ignition mechanisms, the explosive power of cryogenic H2/Ox mixtures under different conditions, and to elucidate the puzzling source of the initial formation of flames near the intertank section during the Challenger disaster. We carry out a physics-based analysis of general explosions scenarios for cryogenic gaseous H2/Ox mixtures and determine their realizability conditions, using the well-established simplified models from the detonation and deflagration theory. We study the features of aerosol H2/Ox mixture combustion and show, in particular, that aerosols intensify the deflagration flames and can induce detonation for any ignition mechanism. We propose a cavitation-induced mechanism of self-ignition of cryogenic H2/Ox mixtures that may be realized when gaseous H2 and Ox flows are mixed with a liquid Ox turbulent stream, as occurred in all HOVI tests. We present an overview of the HOVI tests to make conclusion on the risk of strong explosions in possible liquid rocket incidents and**

[1] Senior Research Scientist, MCT Inc, NASA Ames Research Center, Moffett Field, CA, 94035.
[2] Associate Professor, Department of Mathematical Sciences, New Jersey Institute of Technology, Newark, NJ 07102.
[3] Research Scientist, SGT, Inc., NASA Ames Research Center, Moffett Field, CA, 94035, USA.
[4] Research Scientist, SGT, Inc., NASA Ames Research Center, Moffett Field, CA, 94035, USA.
[5] Senior Research Scientist, Applied Physics Group Lead, NASA Ames Research Center, MS 269-1, Moffett Field, CA, 94035.
[6] Aerospace Engineer, Supercomputing Division, AIAA Member, NASA Marshall Space Flight Center, Huntsville, Alabama.
[7] Aerospace Engineer, AUS, AIAA Member, NASA Marshall Space Flight Center, Huntsville, Alabama 35812.
[8] Aerospace Engineer, Supercomputing Division, AIAA Member, NASA Marshall Space Flight Center, Huntsville, Alabama 35812.

**provide a semi-quantitative interpretation of the HOVI data based on aerosol combustion. We uncover the most dangerous situations and discuss the foreseeable risks which can arise in space missions and lead to tragic outcomes. Our analysis relates to only unconfined mixtures that are likely to arise as a result of liquid propellant space vehicle incidents.**

## Nomenclature

ρ = gas mass density
P = gas pressure
T = gas temperature
u = gas velocity
E = internal energy
C = sound velocity
$C_V$ = specific heat at constant volume
$C_P$ = specific heat at constant pressure
$\gamma$ = ratio of specific heats; $\gamma = C_p/C_V$
$C_i$ = molar concentration of the *i*-th component
$M_i$ = molar mass of the *i*-th component
*L* = size of the mixed H2/Ox clouds
$L_D$ = thermodiffusion length
$R_0$ = universal gas constant
$R_b$ = bubble radius
*R* = reaction rate
*j* = mass flux
$Q_h$ = heat of combustion
*S* = cross-section area
*GH2* = gaseous hydrogen
*LH2* = liquid hydrogen
*GOx* = gaseous oxygen
*Lox* = liquid oxygen
$\kappa$ = thermal conductivity
$\mu$ = dynamic viscosity

Subscripts

*g* = gas
*L* = liquid
*V* = vapor
*H* = hole
*t* = tank
*H2* = hydrogen
*Ox* = oxygen
*N2* = nitrogen
*0* = initial state
*mix* = mixture
*max* = maximum
*ign* = ignition
*evap* = evaporation
$c_j$ = *Chapman-Jouguet point*

## I. Introduction

The Challenger disaster of 1986 provoked studies of different risks that can lead to similar catastrophic events related to the use of H2/Ox cryogenic fuels. As was established by the Challenger investigation, the original source of the disaster was freezing of the O-ring in the lower section of the left solid booster and formation of a gas leak through the O-ring [1-3]. This leak developed into a strong jet of hot gases from the booster and caused the separation of the lower dome from the rest of the LH2 tank. As a result, the tank began to accelerate upwards under the action of the gas pressure inside the tank. The accelerating LH2 tank broke the LOx feed line in the intertank space and then the LOx tank's bottom dome. The resulting LOx stream from the broken LOx feed line was injected into the intertank space, mixing with the GH2 jet from the rupture of the LH2 tank top dome. The mixture of fuels in the intertank space then self-ignited, causing disintegration of the Orbiter and the tragic loss of life (Fig. 1).

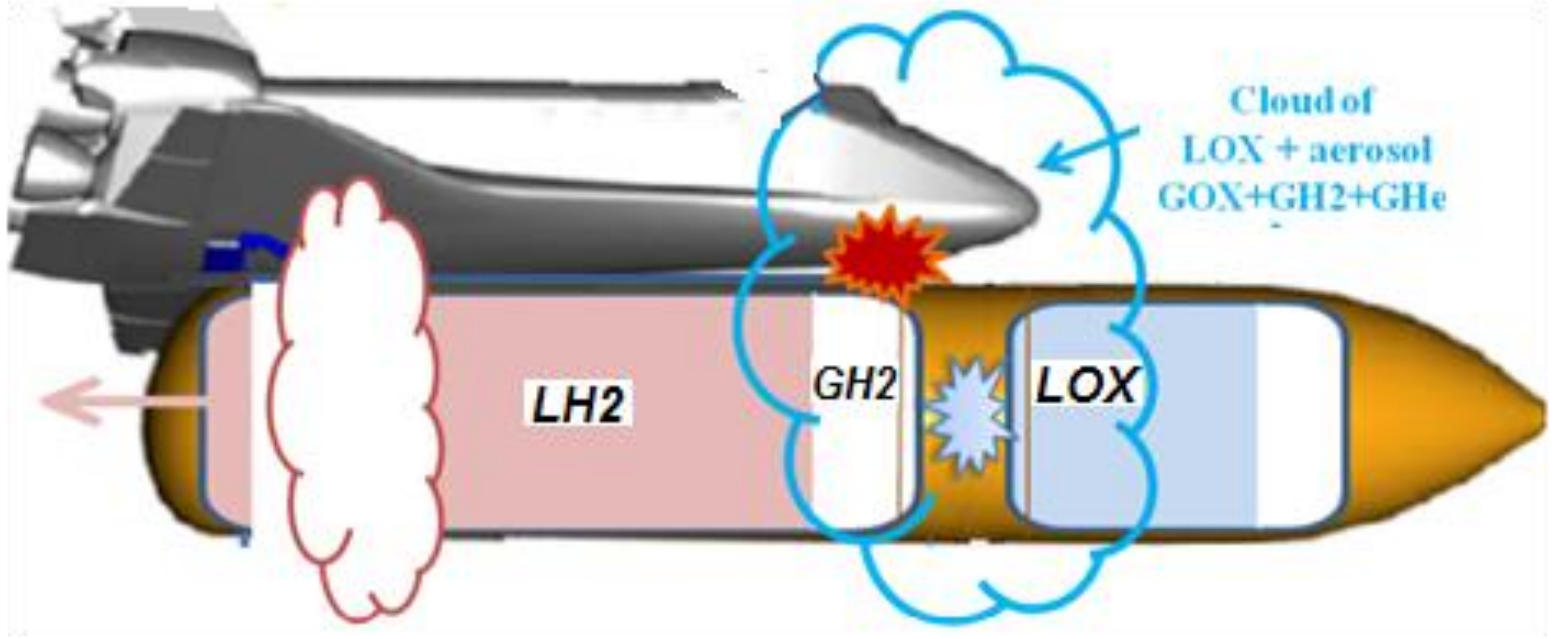

**Fig. 1 Initiation of the first flash near the Challenger's Orbiter/External tank forward attachment.**

The Challenger disaster represents only one possible scenario of a sequence of catastrophic events involving potentially explosive cryogenic fuels such as LH2 and LOx. Another such scenario has to do with an uncontained failure of the first stage of an LH2/LOx-liquid rocket shortly after launch. As a consequence of such an event, the fully loaded tanks of the second and third stages would come crashing down to the ground, violently releasing their entire content into the air. To predict the power of the ensuing explosion, a number of factors determined by the physical processes leading to the explosion have to be taken into consideration. In a situation in which the LH2 tank hits the ground first, the following sequence of events will take place. First, as the LH2 storage tank disintegrates upon impact, LH2 is ejected from the rocket onto the ground. Second, the resulting splash of rapidly evaporating LH2 produces GH2 and a spray of LH2 droplets in the air that are expanding from the impact location. Third, after some delay the rapture of the LOx tank leads to ejection of LOx and the formation of a LOx spray into the GH2-rich area. Direct contact between LH2 and LOx is known to lead to self-ignition of the hydrogen/oxygen gas mixture [4]. The energy released by the LH2/LOx combustion then vaporizes the liquid propellants, increasing their mass in the gas form and makes them available for further reaction, ultimately leading to an explosion.

We note that the analysis of the underlying physical processes and catastrophic risks associated with the use of cryogenic fuels in rocket engines presents a number of puzzles. Up until now, the most enigmatic event in the sequence leading to the Orbiter's disintegration in the Challenger disaster has been the initial formation of flames near the intertank section, not near the engine nozzles [1-3]. In other words, the mixture of cryogenic GH2 and LOx/GOx self-ignited near the intertank region. This is quite surprising, considering the fact that in the tanks LOx is stored at a very low temperature of about 90K and LH2 is stored at an even lower temperature of only about 20K, while the minimum temperature of GH2/GOx mixture self-ignition at atmospheric pressure is about 850K [5]. Similarly, the power of the explosion following an uncontained first stage failure should depend on the degree of H2 and Ox mixing before the ignition occurs. Clearly, only a (possibly small) part of the propellants in which they are well-mixed can participate in the chemical reaction. At the same time, the mixed fraction of the propellants sensitively depends on the time delay between the propellant release and the moment of self-ignition. On the other hand, the character of the explosion (a strong blast or a weak deflagration flame) should strongly depend on the initial density and temperature of the mixed propellants.

To clarify the important questions about the mechanisms governing the outcome of explosions involving cryogens, a set of experiments with LOx and LH2 tanks, the Hydrogen-Oxygen Vertical Impact (HOVI) tests were carried out in NASA Johnston Space Center White Sands Test Facility. We appreciate Dr. Frank Benz for providing us the HOVI test data. The configuration of the tanks in the tests was similar to the one in a launch vehicle, as well as in the External Tank of the Space Shuttle (Fig.2).

In all HOVI tests, the LOx tank was placed above the LH2 tank. The tanks were made from an aluminum alloy and were insulated with 1÷2 inch-thick polyurethane foam. Either the LOx tank alone or both the LH2 and the LOx tank were fixed on a 76 m (250 ft)-high drop tower. Then the tanks were dropped to the ground. The results of these tests showed some surprising characteristics of the explosions involving these cryogenic substances (see Section II for details).

The purpose of this paper is to reconstruct the general picture of different types of cryogenic H2/Ox mixture explosions based on well-know results of combustion theory [5-8] and to present an overview of the HOVI tests to make conclusions about the risk of a strong explosion in possible liquid rocket incidents. We provide a semi-quantitative interpretation of the HOVI data and analyze the ignition conditions and the parameters of different types of cryogenic H2/O2/N2 mixture explosions including the ones involved in the HOVI tests. We carry out a physics-based analysis of the general explosion scenarios and determine their realizability conditions, using the well-established simplified models from the detonation and deflagration theory. We uncover the most dangerous situations and discuss the foreseeable risks which can arise in space missions and lead to tragic outcomes.

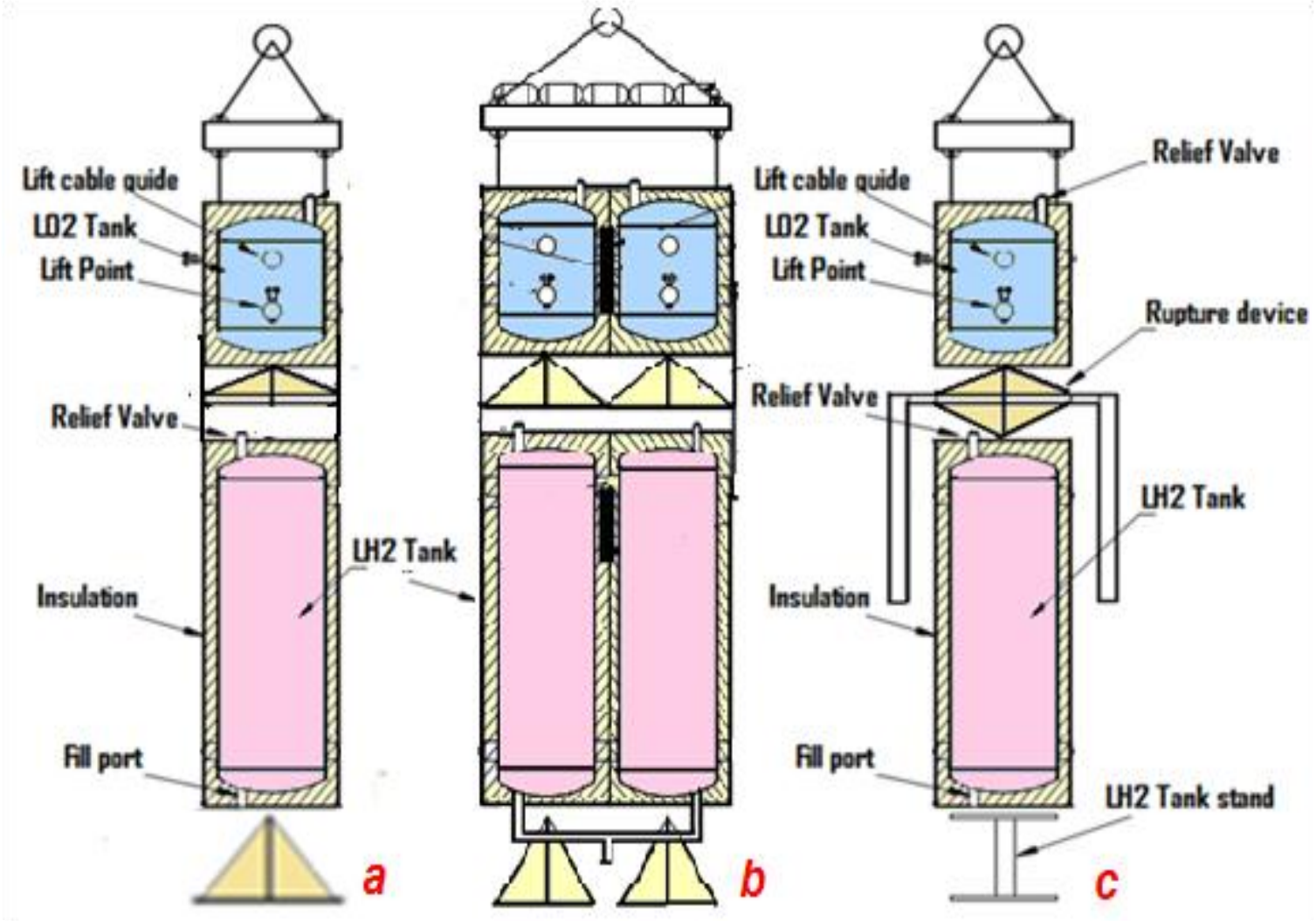

**Fig. 2 Two main types of LH2 and LOx tanks used in the HOVI tests. The first type: two rupture devices are located under the bottoms of both tanks (a) and (b); the second type: one rupture device is between the LH2 and the LOx tank (c).**

The purpose of this paper is to reconstruct the general picture of different types of cryogenic H2/Ox mixture explosions based on well-know results of combustion theory [5-8] and to present an overview of the HOVI tests to make conclusions about the risk of a strong explosion in possible liquid rocket incidents. We provide a semi-quantitative interpretation of the HOVI data and analyze the ignition conditions and the parameters of different types of cryogenic H2/O2/N2 mixture explosions including the ones involved in the HOVI tests. We carry out a physics-based analysis of the general explosion scenarios and determine their realizability conditions, using the well-established simplified models from the detonation and deflagration theory. We uncover the most dangerous situations and discuss the foreseeable risks which can arise in space missions and lead to tragic outcomes.

We note that our analysis is limited to unconfined mixtures that are likely to arise as a result of liquid propellant space vehicle incidents. An independent problem is the H2/O2 mixture explosion in confined spaces, e.g. in the propellant feed lines. In such cases, detonation may arise as a result of the interaction of a deflagration flame with an external shock wave or localized obstacles [6, 9]. These effects are not analyzed in the present paper.

Our paper is organized as follows. The main data for the HOVI tests are analyzed in Section II. The theoretic basis, the conditions and the parameters of detonation, deflagration, and aerosol combustion in cryogenic H2/Ox/N2 mixtures, and also the interpretation of the HOVI test data are presented in Section III. The mechanism of cavitation-induced ignition of these mixtures is considered in Section IV. The conditions, parameters, and risk of the strong blast onset in cryogenic H2/Ox/N2 mixtures are discussed in Section V.

## II. HOVI Test Data: Conditions and Typical Scenarios of Explosions

Most HOVI tests used two tanks, one LH2 tank and one LOx tank (Fig.2a and 2b), except for HOVI 9 and 10 tests, in which the side-by-side double tank configuration consisting of four tanks were used (Fig.2c). The LOx and LH2 tanks in most tests, including HOVI 9, 13, and 14, were fixed on a 76 m (250 ft)-high drop tower. Then both tanks were dropped to the ground. In HOVI 2 and 5 only the LOx tank was dropped on the LH2 tank situated on the ground. The impact velocity was within 30÷35m/sec. All HOVI tests can be divided into two groups based on the location of the rupture devices. The group 1 consists of the HOVI tests in which two rupture devices located under the bottoms of the LOx and the LH2 tank were used (see HOVI 13 and 9 in Fig.2a and b). Group 2 consists of the tests in which only one rupture device located between the LOx and the LH2 tank was used. HOVI 2 and 5 (Fig.2c) belong to this group.

The tanks used had three typical sizes. The LH2 (LOx) tanks in HOVI 13 and 14 have diameter $D_t$=0.94 m, height $H_t$ = 3.84 (1.42) m, the initial fuel mass of about $m_{H2}$=129 (840) kg, the total volume $V_t$=2.396 (0.817) m$^3$ and the gas (ullage) volume $V_{g0}$ =0.553 (0.081) m$^3$. The pressures in the tanks are $p_{H2}$= 1.43 atm and $p_{Ox}$=3.15 atm, respectively. HOVI 9 and 10 have double tanks, with each of the LH2 (LOx) tanks having $D_t$=0.46 m, $H_t$ = 1.78 (0.71) m, $m_{H2}$=33 (176) kg, $V_t$ = 0.273 (0.095) m$^3$, $V_{g0}$= 0.037 (0.018) m$^3$, $p_{H2}$=1.43 atm ($p_{Ox}$=3.42 atm), respectively. The LOx and LH2 tanks in HOVI 2 and 5 have $D_t$=0.58 m, $H_t$ = 2.24 (0.86) m, $m_{H2}$ =37 (189) kg, $V_t$ = 0.545 (0.185) m$^3$, $V_{g0}$ = 0.016 (0.019) m$^3$, $p_{H2}$=1.43 atm ($p_{Ox}$=3.38 atm in HOVI 2 and $p_{Ox}$=5 atm in HOVI 5), respectively.

Many pressure sensors and also three film and video cameras were used to detect the explosion parameters. The main pressure sensors were located along three legs at ground level, 10 sensors in each (Fig.3). The main purpose of these tests was to obtain explosion data that would be more typical or more representative of a launch vehicle failure than the distributive mixture tests.

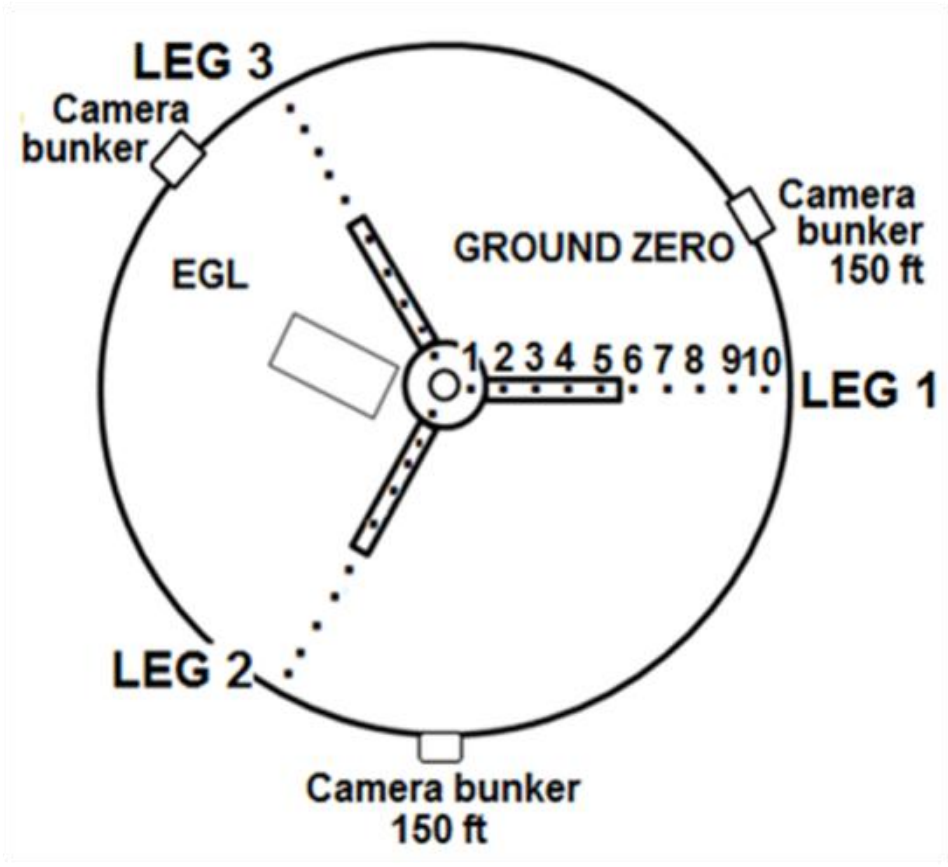

| sensor | m |
|---|---|
| 1 | 1.2 |
| 2 | 1.7 |
| 3 | 2.5 |
| 4 | 3.6 |
| 5 | 5.4 |
| 6 | 7.7 |
| 7 | 11.5 |
| 8 | 16.4 |
| 9 | 23.8 |
| 10 | 35.5 |

**Fig. 3 The HOVI test site and the location of the main pressure sensors.**

Both the HOVI and the LH2/LO2 pan (dewar) tests demonstrated that the ignition always occurred and was not due to external sources. The HOVI test data also showed that a liquid hydrogen spill alone is not likely to self-ignite, because in every HOVI test with a ground cloud of hydrogen, caused by a rupture in the bottom of the hydrogen tank, the ground cloud did not ignite until liquid oxygen was released. The HOVI data verified the tendency for self-ignition of H2/Ox mixtures, because each HOVI test ignited without external assistance. HOVI test data showed that self-ignition occurs when GH2, GOx, and LOx mixture is available.

In each of the test in the first group (HOVI 9, 13, 14) the LOx and the LH2 tanks were raised together. After the impact with the ground, first the lower rupture device (Fig. 2) broke the bottom dome of the LH2 tank, then the upper rupture device broke the LOx tank’s bottom dome with some delay. As a result, H2 and Ox liquid streams were ejected from the tanks. These streams were fragmented and the liquid droplets were partially evaporated. The hydrogen aerosol cloud arose near the LH2 tank bottom area (Fig.4).

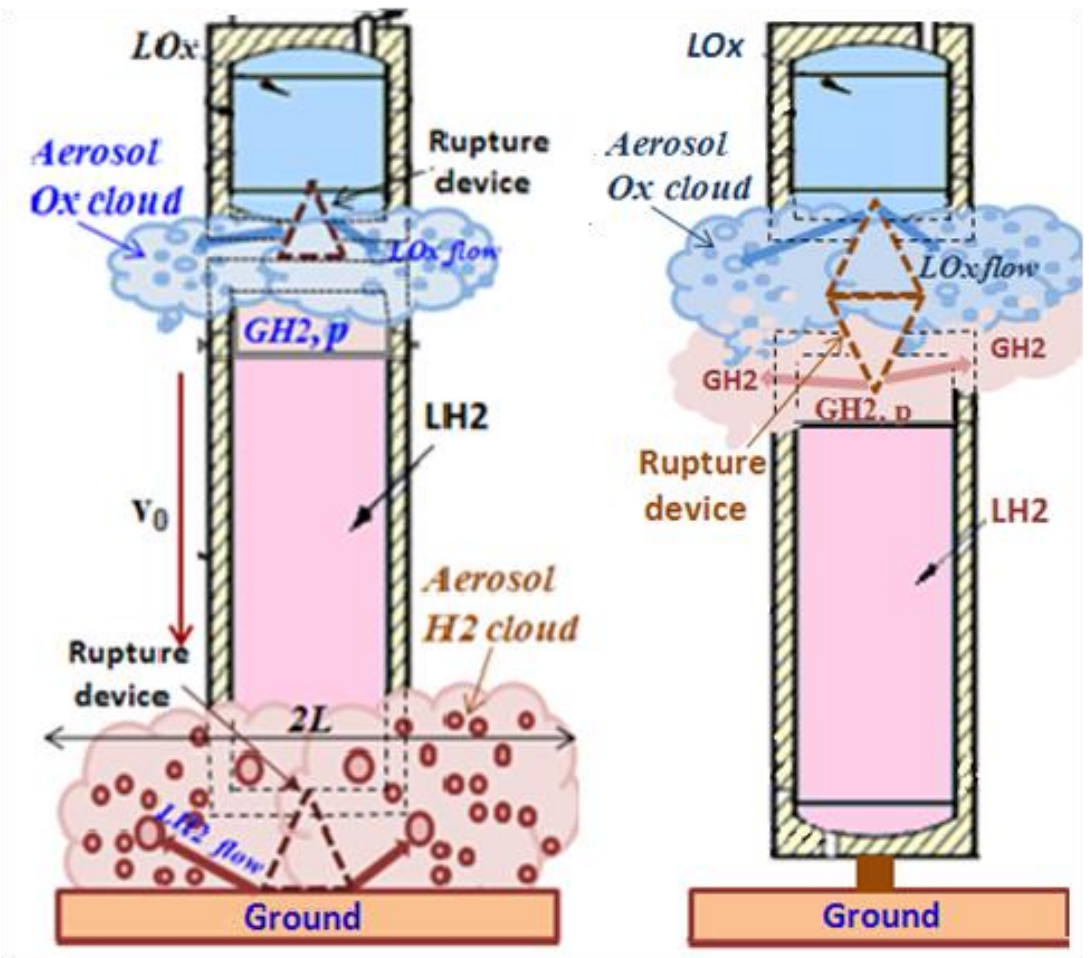

**Fig. 4 Dynamics of the formation of H2 and Ox aerosol clouds and the explosion for HOVI tests of the first (left) and second (right) types.**

The oxygen aerosol cloud appeared in the region between the tanks with the delay time $t_{delay}$*~20-30msec*. Then these aerosol clouds mixed partly and the explosion was observed with the total delay $t_{delay}$*~100msec* that coincides with the time of the fall of LOx pieces onto the ground. The size of the observed clouds was about 2x1.4x3 $m^3$ when the explosions arose. These explosions are characterized by the sensor data presented in Fig. 5. These data are discussed later on.

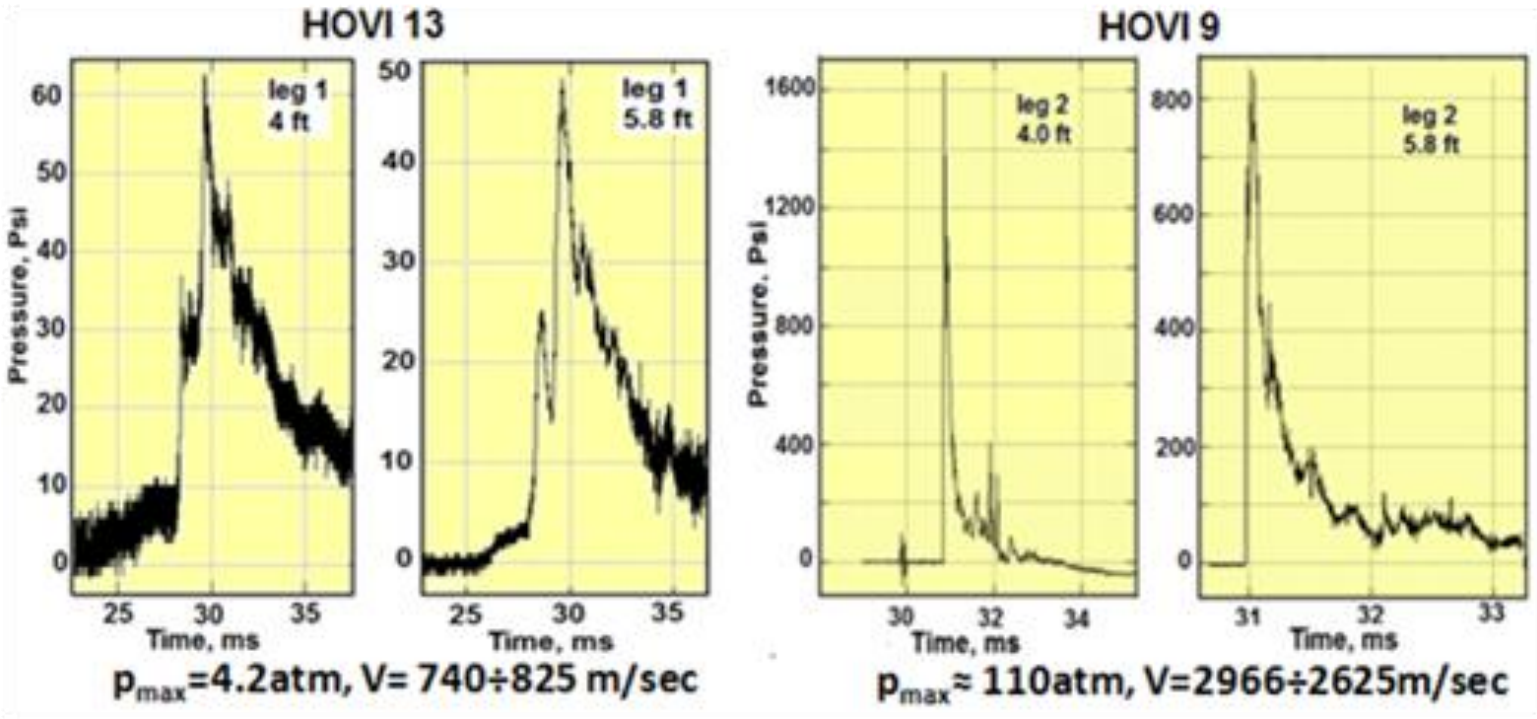

**Fig. 5 Typical data of pressure sensors located near the explosion center for the HOVI tests of the first type. Data are provided by NASA Johnston Space Center White Sand Test Facility.**

In the tests of the second group (HOVI 2 and 5) the LOx tank was dropped onto the LH2 tank that was positioned at ground level. After the impact of the LOx tank with rupture device located between the tanks (Fig. 2c) the rupture device broke the top dome of the LH2 tank and, with some delay, the bottom dome of the LOx tank. As a result, GH2 and LOx streams escaped from the tanks into the region between the tanks (Fig. 6).

The LOx stream was fragmented and the liquid droplets were evaporated partly. The oxygen aerosol cloud appeared with the delay time $t_{delay}$~*20-30msec*. The gaseous hydrogen cloud mixed partly with the LOx aerosol cloud and the explosion was observed with the total delay $t_{delay}$~60ms. These explosions are characterized by the sensor data presented in Fig. 7.

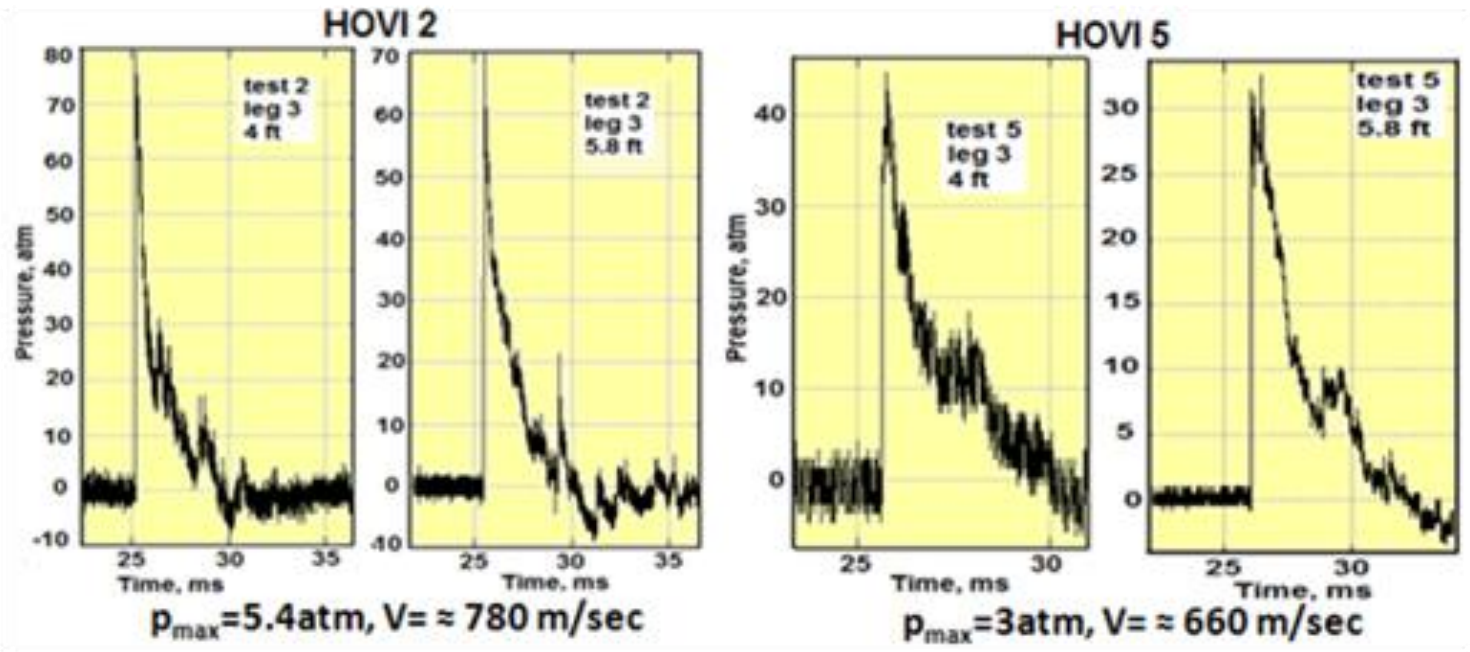

**Fig. 6 Typical data from the pressure sensors located near the explosion center for the HOVI tests of the second type. Data are provided by NASA Johnston Space Center White Sand Test Facility.**

## III. Explosions of Cryogenic H2/O2/N2 Mixtures

To analyze the experimental data and to estimate the risks of a strong explosion, we first consider detonation and deflagration characteristics of cryogenic GH2/GOx/GN2 mixture explosions in unconfined areas as functions of their temperature and composition, including the conditions for initiation of these explosions. Then, we compare these detonation and deflagration characteristics with the HOVI test data.

### A. Stationary Detonation Wave in GH2/GO2/GN2 Mixtures

Detonation is supersonic combustion induced by a strong shock wave propagating directly ahead of the combustion front. The stationary detonation wave in unconfined reactive gas mixtures is described by the Chapman–Jouguet theory extended by Zeldovich, Von Neumann and Doering (ZND) [6-9]. The ZND theory is based on the equations of mass, momentum and energy conservation, the ideal gas equation of state, and the equations describing the kinetics of chemical chain reactions. The ZND model with full chemical kinetics including 21 chemical reactions (*CANTERA* [6]) was used to calculate the characteristics of the detonation waves in different GH2/GO2/GN2 mixtures.

The structure of the detonation wave in the GH2/GO2 stoichiometric mixture and the ZND method of its construction are shown in Fig. 8, where the Hugoniot curve is given by

$$\frac{\gamma}{\gamma-1}\left(\frac{p}{\rho}-\frac{p_0}{\rho_0}\right)-\frac{p-p_0}{2}\left(\frac{1}{\rho}+\frac{1}{\rho_0}\right)=Q, \tag{1}$$

and the Rayleigh line, which is tangent to the Hugoniot curve at the Chapman-Jouguet point. Here $T$, $p$, $\rho$ are the temperature, pressure and mass density of the combustion products after the detonation wave front, and the subscript

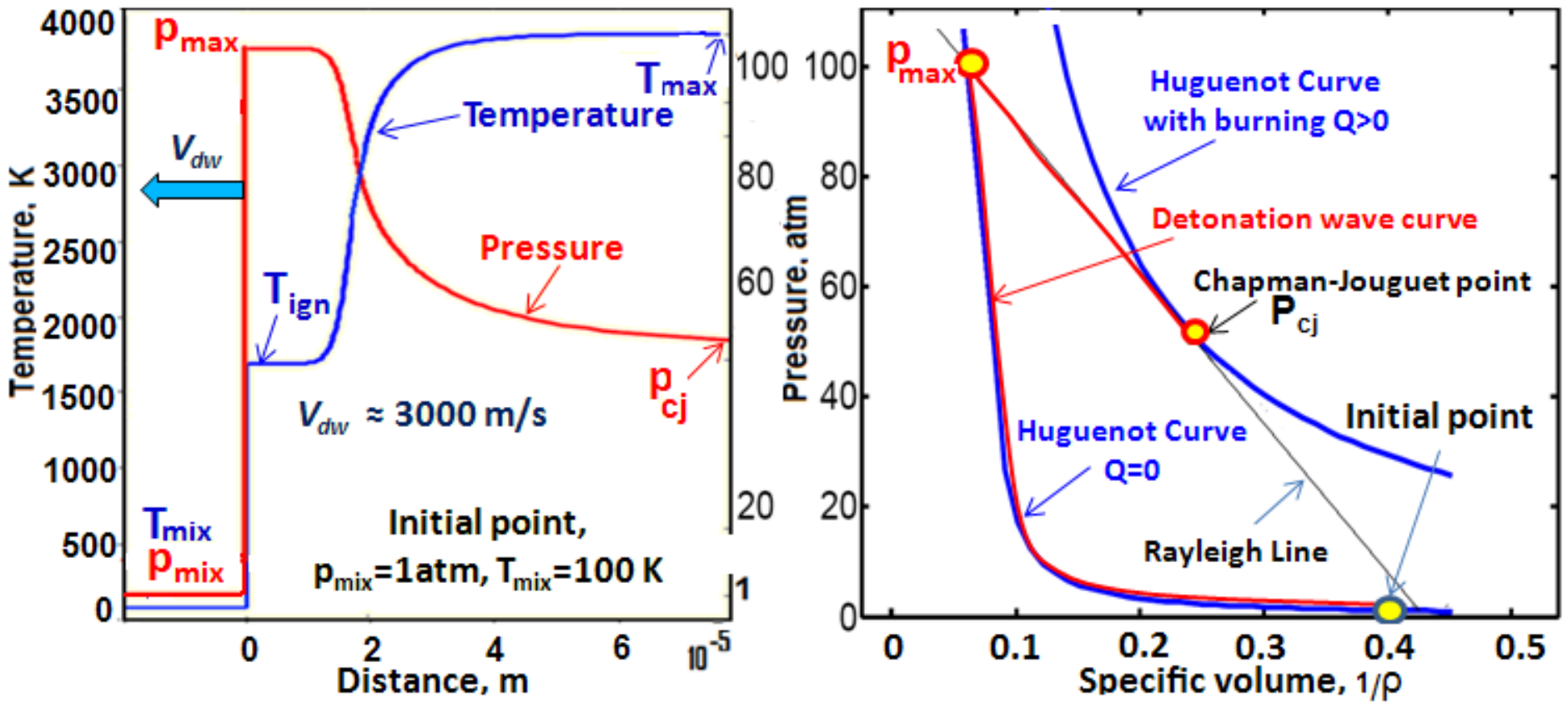

**Fig. 7 A stationary detonation wave in the H2/O2 stoichiometric mixture (2:1) (ZND theory).**

"mix" denotes an initial state of the mixture before the wave front; $Q=\sum Y_i h_i(T)-\sum Y_j^{mix} h_j^{mix}(T_{mix})$ is the reaction enthalpy, $h_i$ and $Y_i$ are the enthalpy and the mole fraction of the $i$-th component of the mixture.

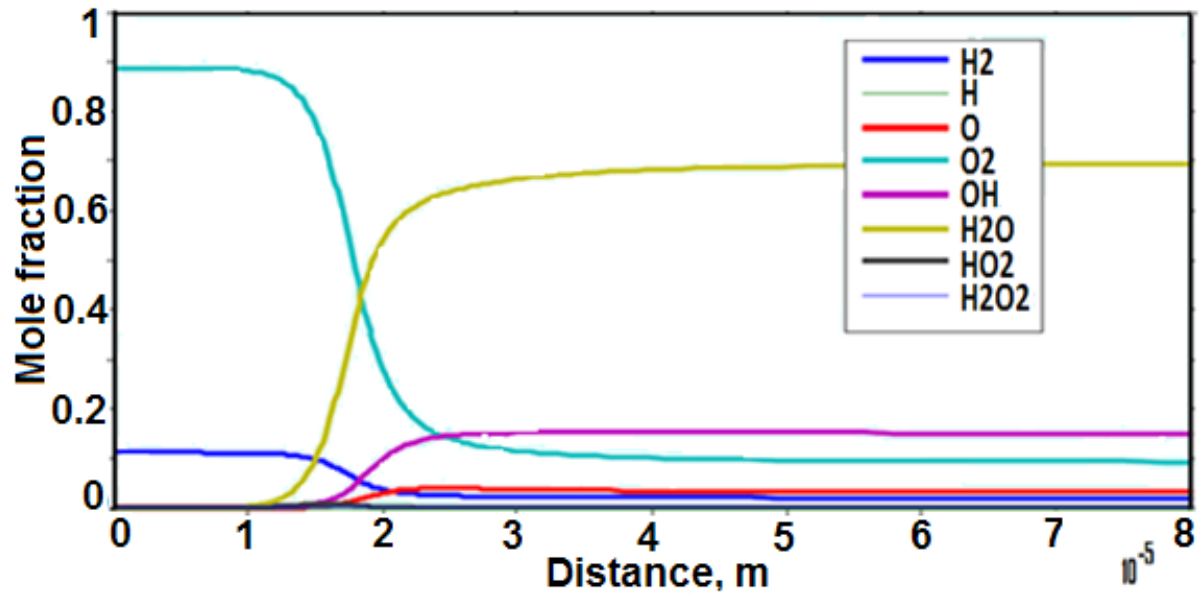

**Fig. 8 A stationary detonation wave in the H2/O2 stoichiometric mixture (2:1) (ZND theory).**

The characteristics of the detonation waves for several GH2/GO2/GN2 mixtures are shown in Fig. 9. Our simulations showed that, as expected, the maximum temperature $T_{max}$ and velocity $v_{dw}$ of the detonation wave strongly depend on the composition and weakly on the initial mixture temperature $T_{mix}$. Conversely, the maximum detonation pressure $p_{max}$ depends strongly on $T_{mix}$ and relatively weakly on the mixture composition (Fig. 9). The maximum pressure behaves approximately as $p_{max} \propto (1/T_{mix})^n$, where $n \approx 1.7$. The closer the mixture composition to the stoichiometric H2/O2 composition (2:1), the higher the pressure and temperature of the detonation wave. The highest pressure (≈110atm) and temperature (≈3900K) in the detonation wave, i.e. the strongest blast, are achieved in cryogenic stoichiometric H2/O2 mixtures (2:1) at temperatures $T_{mix} \leq$100K. The blast power is determined by the explosion impulse $p_{max}\tau_{imp}$, where $\tau_{imp}=L/v_{dw}$ is the blast impulse duration, $L$ is the size of the mixed clouds, i.e. the radius of a hemispherical area where the mixed clouds are localized. For example, $p_{max}\tau_{imp} \approx 10^4$ *Pa·sec* for the explosion of the stoichiometric H2/O2 mixture localized in a hemisphere of radius 2.8m. We note that the detonation wave front is unstable with respect to the formation of an irregular cellular structure [6, 7]. However, the main parameters of real detonation waves are close to those given by the ZND theory [13, 14].

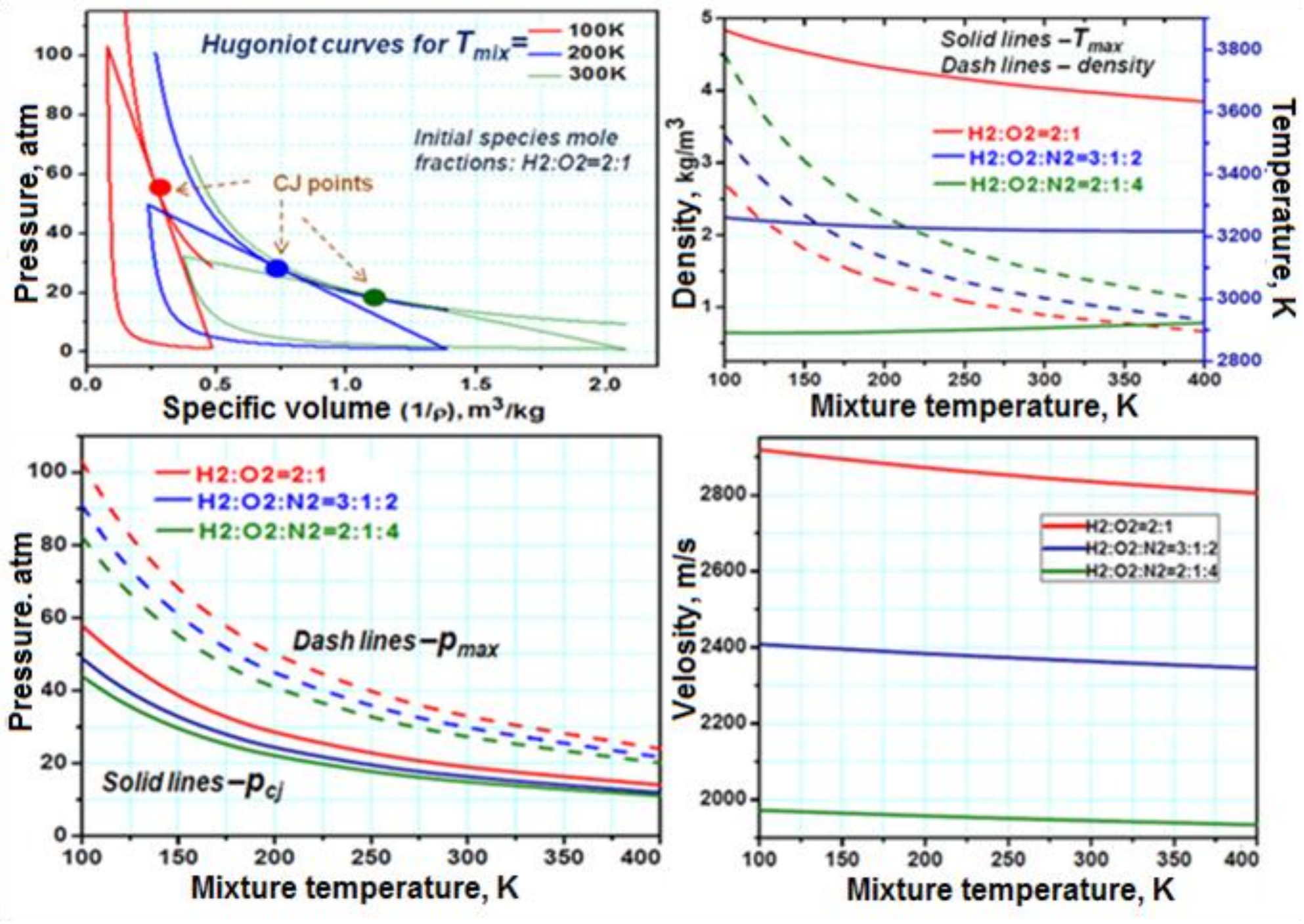

**Fig. 9 The parameters of stationary detonation waves in typical H2/O2 /N2 mixtures.**

### B. Conditions and Dynamics of Detonation Formation in Cryogenic Unconfined Mixtures

One of the possible detonation initiation mechanisms was pointed out by Zeldovich [6-9]. It can be realized only in sufficiently hot explosive mixtures in the presence of a small temperature gradient when the ignition phase velocity in the adjacent parts of the burning mixture is higher than the detonation wave velocity $v_{dw}$. This mechanism and its realization in experiments are analyzed in detail in [10, 11]. At cryogenic temperatures, however, the Zeldovich mechanism is not realized, since the gas mixture is too cold.

Obviously, the detonation of cryogenic GH2/GO2/GN2 mixtures can be induced by a strong external shock wave with the pressure $p > p_{max}$. Such a shock wave can be generated by a local blast. Recent experiments show that detonation in the stoichiometric hydrogen/air mixture enclosed in a large hemispherical envelope can be initiated by a blast of 10 g of C-4 high explosive located in the center of the hemisphere [12]. The detonation velocity was found to be 1980 m/s, which is in good agreement with the results of the ZND theory for the stoichiometric mixture.

To analyze the possible scenarios and conditions of detonation initiation, we will use a simplified combustion model taking into account that the GH2/GOx burning rate is mainly limited by the initiation reactions H2+Ox→OH+OH and H2+Ox→HOx+H, which have the lowest rate. The rate of these reactions can be written as $R_{burn} = A(T)C_{H2}C_{Ox} \equiv C_{H2}/\tau_{H2}$ where $A(T) = 5.5\cdot 10^{7}\exp(-\frac{19680}{T}) + 0.74T^{2.43}\exp(-\frac{26926}{T})\,[m^3/mol\cdot s\mathrm{ec}]$ [14] and $\tau_{H2}$ is the time scale of $C_{H2}$ variation. Note that this approximation is close to the one-step mechanism of Mitani and Williams [15]. In this case the dynamics of the GH2/GO2/GN2 mixture combustion are described by the continuity equations for the molar concentration of the mixture components $C_i$:

$$\frac{\partial C_{H2}}{\partial t} + \nabla\cdot\left(uC_{H2}\right) = -2C_{H2}C_{Ox}A(T), \quad \frac{\partial C_{Ox}}{\partial t} + \nabla\cdot\left(uC_{Ox}\right) = -C_{H2}C_{Ox}A(T), \tag{2}$$

$$\frac{\partial C_{N2}}{\partial t} + \nabla\cdot\left(uC_{N2}\right) = 0, \quad \frac{\partial C_{H2O}}{\partial t} + \nabla\cdot\left(uC_{H2O}\right) = 2C_{H2}C_{Ox}A(T) \tag{3}$$

as well as conservation of momentum and energy

$$\rho\frac{\partial u}{\partial t} + \rho u\cdot\nabla u = -\nabla p, \quad \rho = \sum_i M_i C_i, \quad p = R_o CT \tag{4}$$

$$\frac{\partial E}{\partial t} + \nabla\cdot[u(E+p)] = \nabla\cdot\left(\kappa\nabla T\right) + 2Q_h C_{H_2}C_{O_2}A(T), \quad T = \frac{E-\rho u^2/2}{5CR_0/2}. \tag{5}$$

Here $C = \sum_i C_i$, $M_i$ and $\kappa_i$ are the molar mass and the thermal conductivity of the *i*-th component. For the analysis we use the following values: $R_0 = 8.31 \dfrac{J}{mol \cdot K}$, $Q_h = 2.86 \cdot 10^5 \dfrac{J}{mol}$, $\kappa = \left( \dfrac{T}{T_{ref}} \right)^{1/2} \sum_i \dfrac{\kappa_i C_i}{C}$, where $T_{ref}$ is a reference temperature. We assume that a radially symmetric explosive mixture with temperature $T_{mix}=100K$ at pressure $p=p_{atm}$ occupies a hemisphere of radius $L_{mix}$ and a source of ignition is in its center. Our simulations show that the detonation arises when the ignition source generates a shock wave with $p_0$=80atm >$p_{c\,j}$=55atm inside a small area of radius $r < r_0$=1cm (Fig. 10).

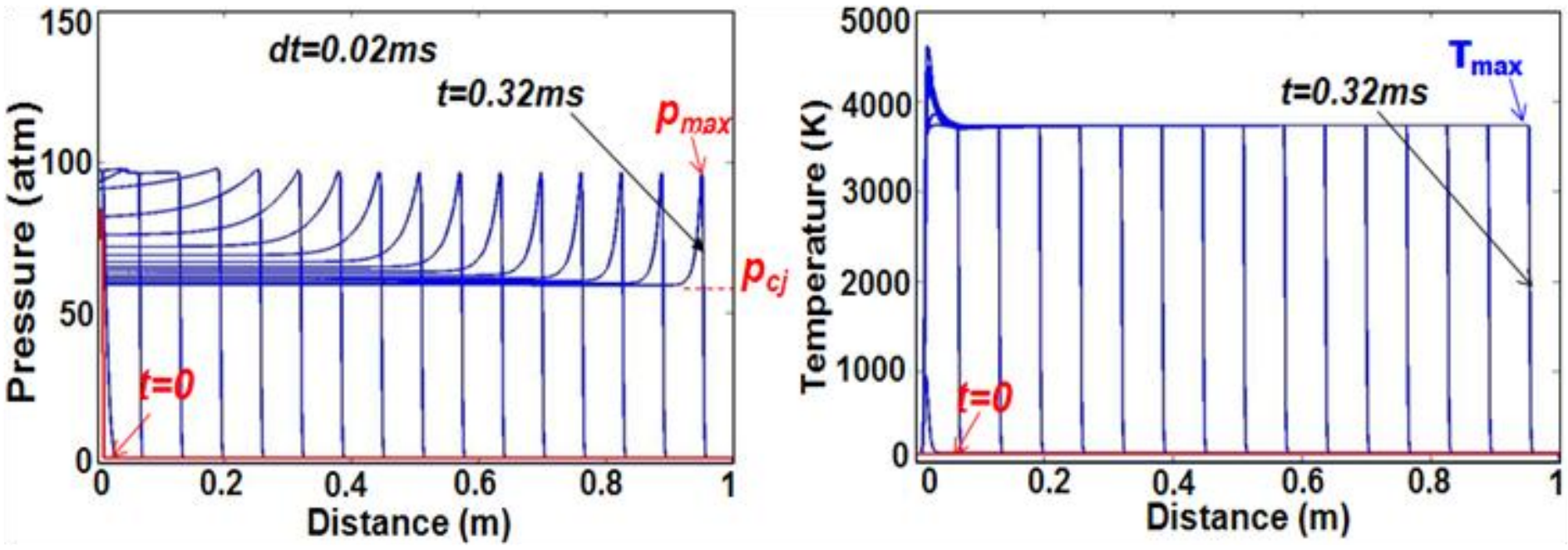

**Fig. 10 Formation of a detonation wave in the stoichiometric GH2/GOx mixture. Initial conditions: $p_0$=1atm and $T_0$=100K for $r$> $r_0$=1cm and $p_0$=80atm and $T_0$=100K for $r < r_0$=1cm. The detonation wave parameters: $p_{max}$=100atm, $p_{cj}$=60atm, $T_{max}$=3800K, $v_{dw}$=3000m/s.**

Parameters of detonation wave obtained: $p_{max}$=100atm, $T_{max}$=3800K and velocity $v_{dw}$=3000m/s are close to those of a stationary detonation wave estimated by the ZND model (Fig. 8). If the initial pressure $p_0$=40atm < $p_{cj}$=55atm inside a small area of radius $r < r_0$=1cm, the shock wave dissipates and no combustion is initiated (Fig. 11).

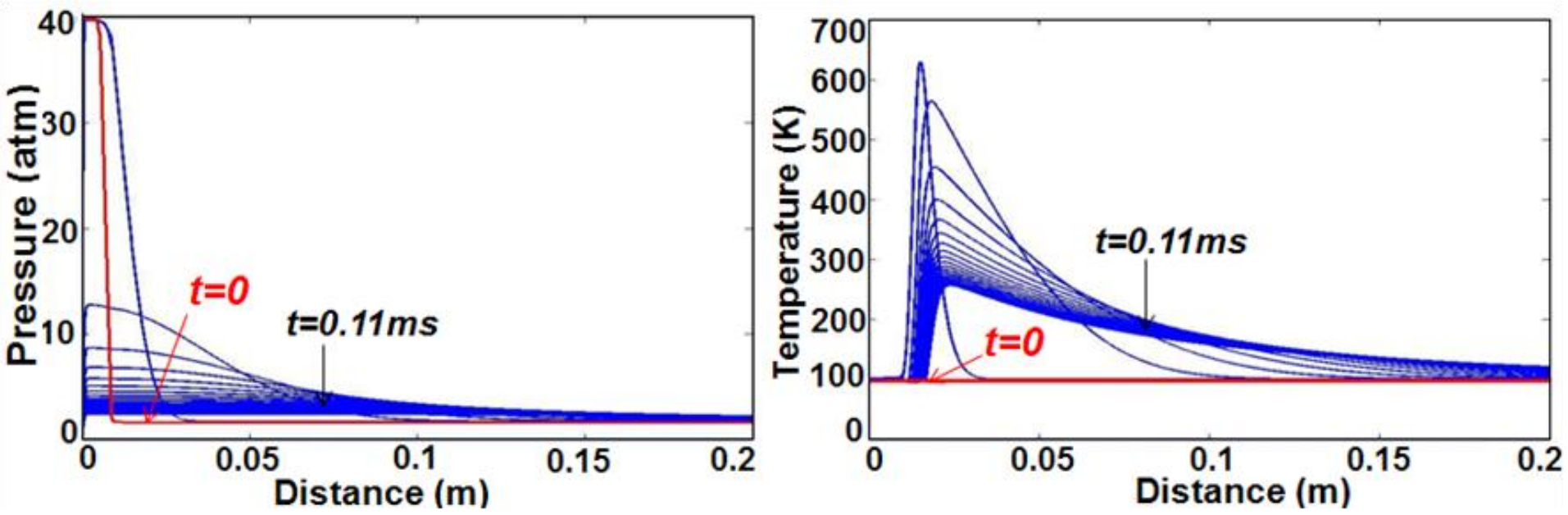

**Fig. 11 The distributions of pressure and temperature in the stoichiometric GH2/GOx mixture. Initial conditions: $p_0$=1atm and $T_0$=100K for $r$> $r_0$=1cm and $p_0$=40atm and $T_0$=100K for $r < r_0$=1cm: no detonation.**

When the initiating shock wave has high enough temperature, $T>T_{ign,}$ then detonation arises even at $p_0$=40atm < $p_{cj}$ (Fig. 12). In this case the parameters of the detonation wave are the same as those for the case presented in Fig. 5.

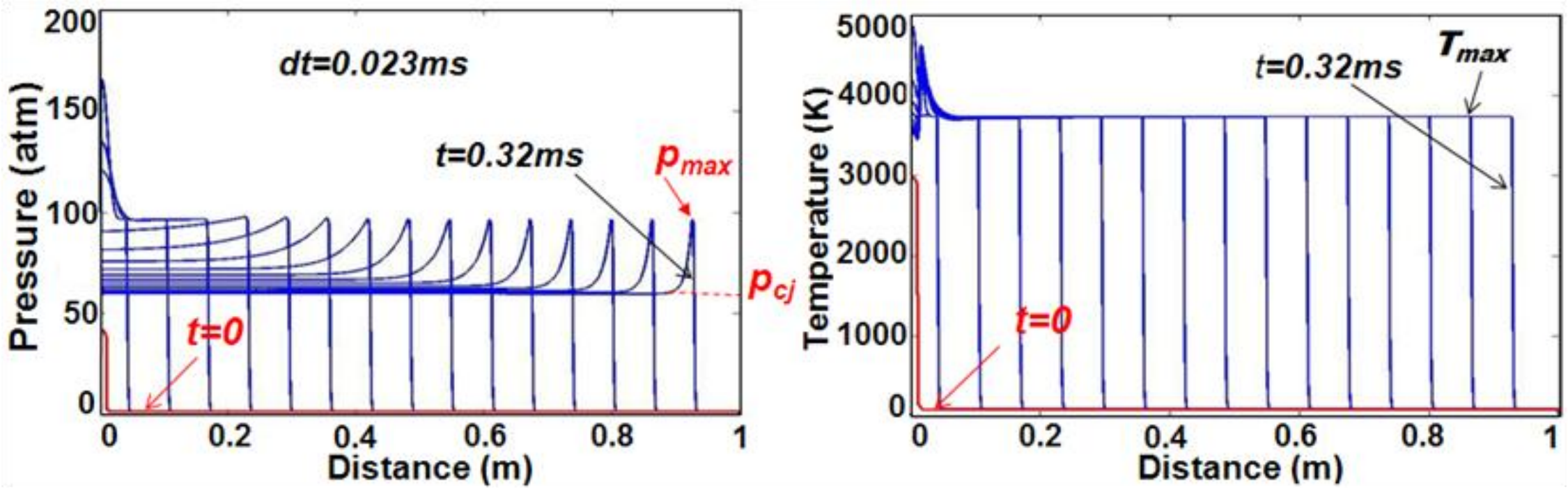

**Fig. 12 Formation of a detonation wave in the stoichiometric GH2/GOx mixture. Initial conditions: $p_0$=1atm and $T_0$=100K for $r> r_0$=1cm and $p_0$=40atm and $T_0$=3000K for $r < r_0$=1cm. The detonation wave parameters: $p_{max}$=100atm, $p_{cj}$=60atm, $T_{max}$=3800K, $v_{dw}$=3000m/s.**

We note that the detonation of the GH2/GO2/GN2 mixture at room temperature is initiated easier than in the cryogenic mixture, since the pressure of the initiation shock wave can be less (Fig. 13). At the same time, the pressure and temperature of the detonation wave in GH2/GO2/GN2 mixtures are smaller than those for the stoichiometric mixture. More general conditions for detonation of the GH2/GO2/GN2 mixtures are discussed in next sections.

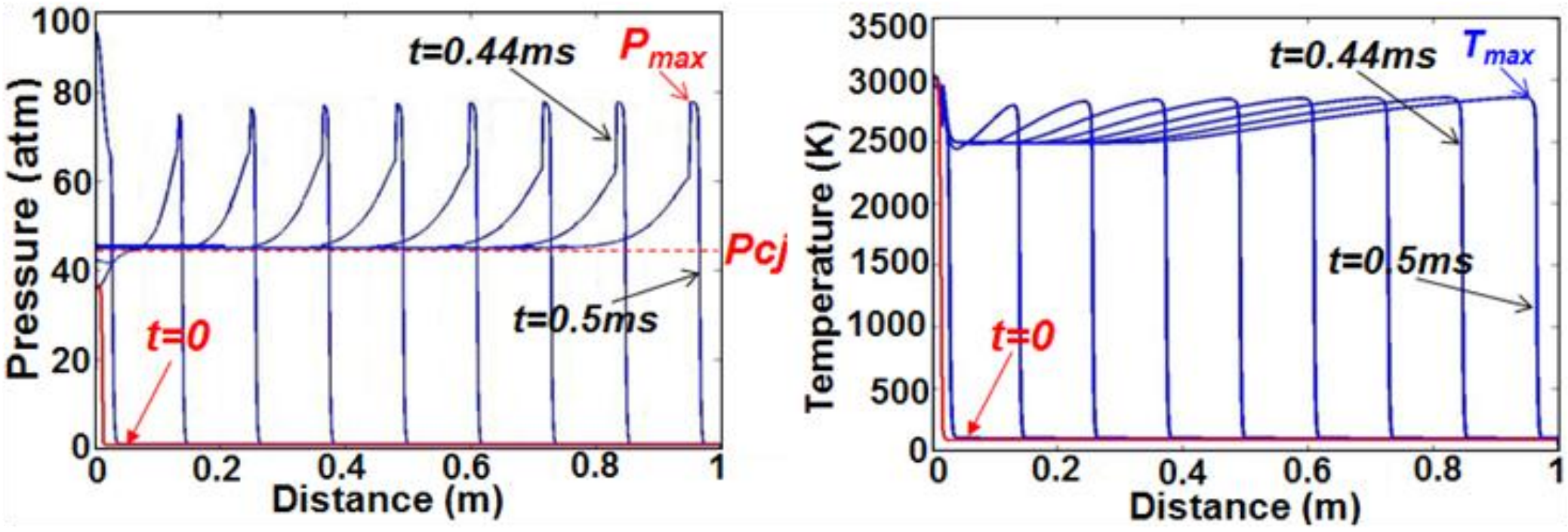

**Fig. 13 Formation of a detonation wave in the GH2/GOx/GN2 mixture (2:1:4). Initial conditions: $p_0$=1atm and $T_0$=100K for $r> r_0$=1cm and $p_0$=35atm and $T_0$=3000K for $r < r_0$=1cm. The detonation wave parameters: $p_{max}$=80atm, $p_{cj}$=45atm, $T_{max}$=2800K, $v_{dw}$=2000m/s.**

### C. Deflagration in unconfined GH2/GO2/GN2 mixtures

Deflagration is slow subsonic combustion mediated by heat conduction: the hot burning gas heats the adjacent layer of the cold gas and ignites it. There are three main processes determining the deflagration flame dynamics at pressures close to $p_{atm}$: (i) conductive heat transfer from the flame front to the cold mixture, (ii) burning rate enhancement due to turbulence, and (iii) thermal expansion of hot combustion products. The burning rate due to the conductive heat transfer in a quiescent gas can be estimated as

$$v_D = \frac{L_D}{\tau_{H2}} = \sqrt{\frac{\kappa_{air} R_b}{C_{air}\rho_{air}}} \approx (1.5 \div 2) m/\sec \tag{6}$$

where $\tau_{H2} \approx 3 \cdot 10^{-6} \sec$ is the typical reaction time of the GH2/GO2/GN2 mixtures. Turbulence accelerates the flame front by increasing the effective flame area. According to experimental and numerical studies, the burning rate increases approximately as [16]:

$$v_{Turb} \approx 3.6 v_D \approx (3 \div 5) m/\sec \tag{7}$$

Moreover, the flame front velocity increases due to expansion of the hot combustion products as

$$\begin{aligned} v_{front} &\simeq v_{Turb}\frac{T_{flam}}{T_{mix}} \approx (20 \div 150) m/\sec, \\ T_{flame} &= \frac{\rho_{H2} Q_{H2}}{\rho_{products} C_{p,products}} + T_{atm} \approx (2000 \div 3900) K \end{aligned} \tag{8}$$

for $T_{mix}$ =*100K÷300K*.

Deflagration can be initiated by a small spark or any hot object with temperature $T > T_{ign}$ (Fig. 14).

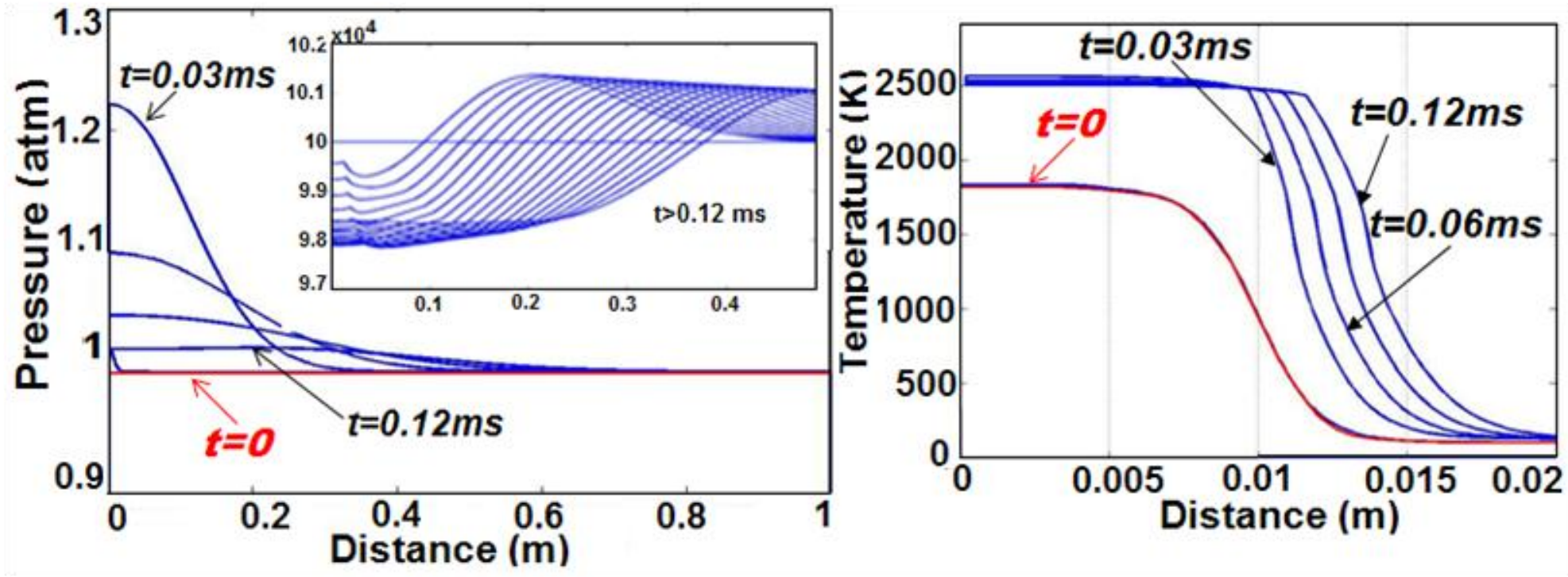

**Fig. 14 Formation of a deflagration wave in the stoichiometric GH2/GOx mixture: the pressure and temperature distributions at different times as a function of distance from the center of the hemisphere combustion wave. Initial conditions: $p_0$=1atm and $T_0$=300K for $r > r_0$=1cm and $p_0$=1atm and $T_0$=1800K for $r < r_0$=1cm (in red). The deflagration wave parameters: $p_{max} \cong$1atm, $T_{max}$=2500K, $v$=28m/s.**

Recent studies of stoichiometric $H_2$/Ox mixture explosions show that in atmosphere the flame front velocity is $v_f = 20 m/sec - 33 m/sec$ and the pressure is close to 1atm [17].

The simulation results of the simplified model given by Eqs. (2)-(5) agree with the results obtained from the analytical estimates above and experiments [17]. The pressure in the deflagration wave is very close to the atmospheric pressure $p_{atm}$. The pressure length scale is much greater than that of temperature, i.e. the “temperature wave” is more localized than the “pressure wave”. This is in contrast to the detonation wave, where the temperature and pressure waves have the same length-scale (see Section A). The pressure in the deflagration wave falls to the atmospheric independently of the initial pressure of the initiating shock wave, if $p < p_{cj}$ (Fig. 15 and Fig. 16).

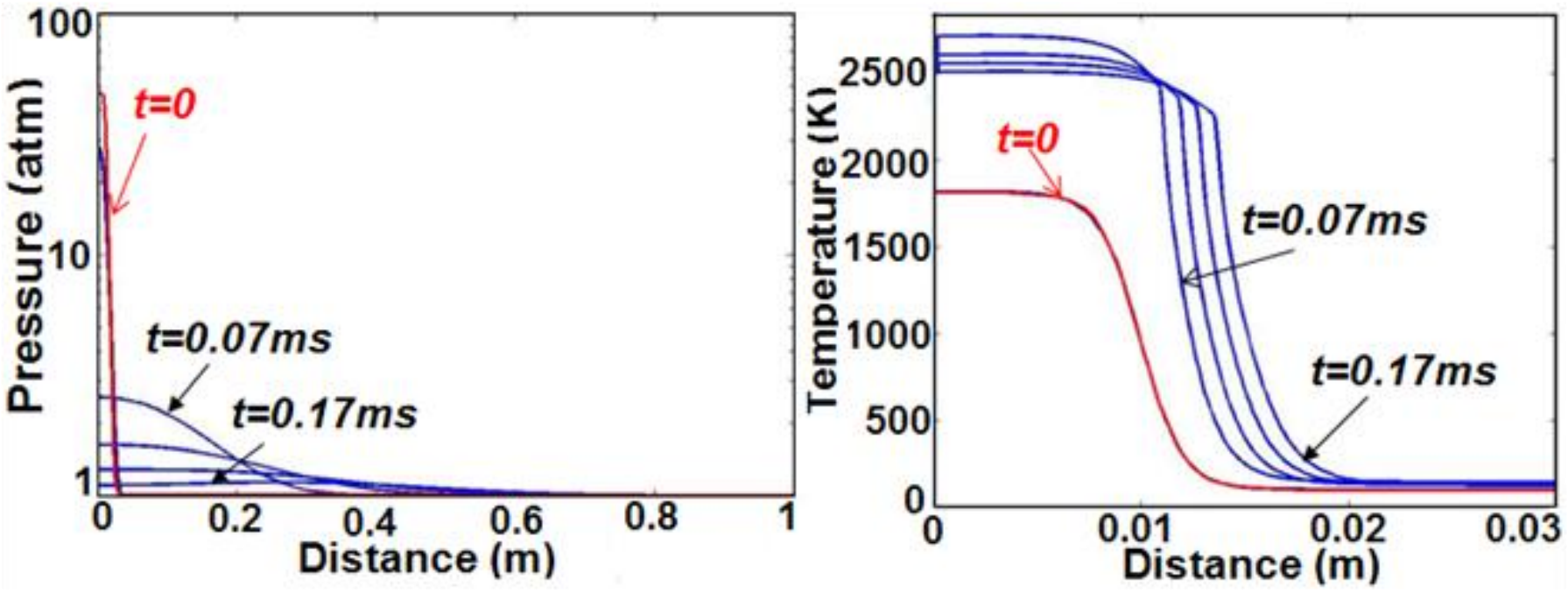

**Fig. 15 Formation of a deflagration wave in the stoichiometric GH2/GOx mixture. Initial conditions: $p_0$=1atm and $T_0$=300K for $r > r_0$=1cm and $p_0$=35atm and $T_0$=1700K for $r < r_0$=1cm (in red). The deflagration wave parameters: $p_{max}$=1atm, $T_{max}$=3000K, $v$=30m/s.**

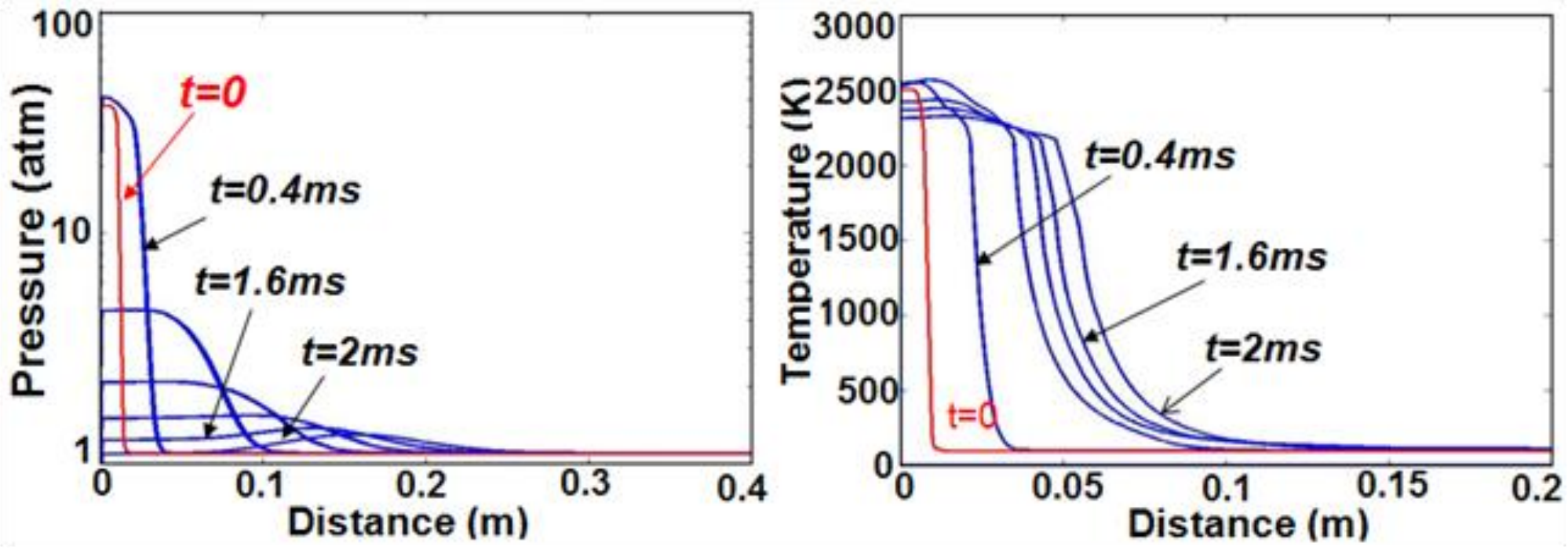

**Fig. 16 Formation of a deflagration wave in the GH2/GOx/N2 mixture (2:1:4). Initial conditions: $p_0$=1atm and $T_0$=300K for $r > r_0$=1cm and $p_0$=35atm and $T_0$=2500K for $r < r_0$=1cm. The deflagration wave parameters: $p_{max} \cong$1atm, $T_{max}$=2200K, v=25m/s.**

**D. Differences between the HOVI test data and the parameters of the detonation and deflagration waves in cryogenic GH2/GO2/GN2 mixtures**

The analysis above shows that the deflagration waves propagating in cryogenic GH2/GO2/GN2 mixtures have pressure $p \approx p_{atm} = 1atm$ and flame velocity $v_{defl} \approx (25 \div 100)\ m/sec$, while the detonation waves have $p > 40atm$ and $v_{dw} > 2000m/sec$. At the same time, the sensor data showed that the maximum pressure in all HOVI tests, except for HOVI 9, were about (3÷5) atm and the front velocities were about (660÷780) m/sec (Fig. 5 and Fig. 7). Thus, these data conflict with the predictions for the detonation and deflagration waves propagating in cryogenic GH2/GO2/GN2 mixtures. We point out that in reality aerosol H2/O2/N2 mixtures, not pure gaseous GH2/GO2/GN2 mixtures, appear in the HOVI tests. The HOVI data may be explained by aerosol combustion of these mixtures, as we discuss below.

**E. Formation of aerosol H2/O2/N2 mixtures in HOVI tests**

Let us first discuss the first HOVI group (Fig. 2 a), in which the impact of the tanks with the rupture devices results in the rupture of the bottom domes in both the LH2 and the LOx tank. The turbulent LH2 jet from the breach impinges on the hot ground and breaks into droplets. The rupture of the bottom dome of the LO2 tank occurs with a delay time $t_{delay} \sim 20\text{-}40msec$ for HOVI 13 and 14. The escaping LO2 stream brakes into droplets after the impact with the LH2 tank top (Fig. 4). The droplets scatter from the tank surface during $\sim 60\ msec$.

Fragmentation of a liquid stream into droplets is a complex and poorly understood phenomenon. Droplet sizes may vary significantly. The typical droplet radius depends on the parameters of both the liquid and gas [18]. Recent experimental studies of liquid jets impinging on a flat smooth surface established the following empirical correlation for the mean droplet radius [19]:

$$r_d = 2.53\times10^5 d_h \,\mathrm{Re}^{-1.28} We^{0.4} \left(\mu_{LH2} / \mu_{air}\right)^{-1.16}, \mathrm{Re} = \frac{d_h v \rho_L}{\mu_{LH2}}, \ \mathrm{We} = \frac{d_h v^2 \rho_L}{\sigma_{LH2}}, \quad (9)$$

Here μ is the dynamic viscosity, which equals to $\mu_{air} = 1.63\times10^{-5} Pa\sec$ for air, $\mu_{LH2} = 1.32\times10^{-5} Pa\sec$ for LH2, and $\mu_{LO2} = 1.96\times10^{-4} Pa\sec$ for LO2, $v$ is liquid velocity, $\rho_L$ is the liquid density, $d_h$ is the stream (hole) diameter, and σ is the surface tension. Assuming $v=v_0=30m/sec$ (see Section II), we find that the typical radius of the droplets in the H2 cloud is about $r_{dr,H2}$= 8mm and in the $O_2$ cloud is about $r_{dr,O2}$= 0.5 mm. We emphasize that these values are almost independent of the hole diameter $d_h$.

The typical droplets of radius $r_{dr}$ and mass $m = 4\pi\rho_L r_{dr}^3 / 3$ bouncing off the ground move with initial velocity of order $v_0$ and fly through the air with temperature $T_{air}$ ≈ 300K and pressure $p_{air}$ = 1atm. The droplet velocity $v$ is slowed by the air drag and is governed by:

$$m\frac{dv}{dt} = -\frac{C_d}{2}\rho_{air}v^2\pi r_{dr}^2 \ or\ \frac{1}{v_0}\frac{dv}{dt} = -\frac{1}{\tau_v}\left(\frac{v}{v_0}\right)^2,\ \tau_v = \frac{8\rho_L r_{dr}}{3\rho_{air}C_d v_0}. \tag{10}$$

Therefore, the velocity and the travel distance as functions of time $t$ are equal to:

$$v(t) = \frac{v_0}{1 + t/\tau_v},\ L_v(t) = v_0\tau_v \ln\left(1 + \frac{t}{\tau_v}\right). \tag{11}$$

Assuming the drag coefficient for a droplet to be $C_d$≈0.4, we find from Eq. (11) that the travel distance for HOVI 13 and 14 is equal to $L_v$≈ 2.7m for the typical LH2 droplets and the explosion delay time $t$ ≈90msec, and $L_v$ ≈ 2.8m for the typical LOx droplets for the explosion delay time $t$ ≈60msec. These values agree with the observed sizes of the H2 and Ox aerosol clouds (Fig. 4 and Fig. 6).

Let us now discuss the second HOVI group, in which the rupture device is between the bottom of the LOx tank and top of the LH2 tank (Fig. 2c and Fig.6). In this case the breach of the LH2 tank top dome results in the escape of gaseous H2, while the breach of the LOx tank bottom results in the escape of liquid Ox. The released LO2 flow is broken into droplets after the impact with the LH2 tank top. Our estimates show that the typical radius of the droplets in the LOx cloud is about $r_{dr,Ox}$=0.5 mm and the scattering distance during the explosion delay time $t$ ≈90msec for HOVI 2 and 5 is equal to $L_v$≈ 2.7m for typical droplets.

### F. Aerosol explosion and the interpretation of the HOVI data

The LH2 and LOx droplets are evaporated partly by the contact with the hot combustion products and air. Therefore, aerosol clouds containing liquid droplets and gaseous H2, Ox, and N2 are formed. During time $\tau_d$ ≈*100msec*, H2 and Ox clouds partly mix and the gaseous H2 and Ox mixture ignites (see Sections A and D). As we noted in Section C, the temperature of such a burning mixture can be about $T_{flame}$*=3000K÷3900K*. To understand the effect of the increased GH2 and GOx masses due to droplet evaporation within the aerosols, we need to pinpoint the primary heat transfer mechanism from the hot combustion products to the LH2 and LOx droplets. We note that this mechanism should be very efficient in order to add a significant amount of extra fuel to the mixture during the short explosion duration of a few milliseconds in order to have a significant amplifying effect.

A natural candidate for the primary heat transfer mechanism is heat conduction from the hot combustion products through the gas phase [6]. Let us estimate the fraction of the droplet mass that would evaporate by heat conduction during time $\tau_d \approx 3m\sec$ corresponding to the typical observed duration of the pressure spike in the aerosol cloud. If $L_D = \sqrt{\kappa_g \tau / \rho_g c_g} \approx 0.5\,\mathrm{mm}$ is the thermodiffusion length in the gas phase, where we took $\kappa_g = \kappa_{g,ref}(T_{flame}/T_{ref})^{1/2} \approx 0.1W/(m \cdot K)$ as the gas thermal conductivity relative to some reference value (for air at room temperature), $\rho_g \approx 1\,kg/m^3$ as the gas mass density, and $c_g \approx 10^3\,J/(kg \cdot K)$ as the specific heat at constant pressure, then the heat balance $m_{evap} q_L \simeq 4\pi r_{dr}^2 \tau_d \kappa_g T_{flame} / L_D$, yields $m_{evap} \approx 4 \times 10^{-3}\,g$ for an LH2 droplet of radius $r_{dr,H2} = 8\;mm$, which weighs $m_{dr} = 4\pi r_{dr}^3 \rho_L / 3 = 0.15\,g$. This estimate shows that only a small fraction (a few per cent) of the droplet mass may evaporate during the short explosion time. A similar situation takes place for LOx droplets of radius $r_{dr,Ox} = 0.5\;mm$. This is due to the fact that in an aerosol formed as a result of a splash the droplets have rather large radii, which prevents them from being evaporated efficiently by the conductive heat transfer mechanism. Other situation occurs for very small spray droplets [6].

Another possible mechanism for heat transfer from the hot combustion products to the droplets is via radiation. We note that in HOVI tests the explosion was accompanied by a strong flash of bright white light similar to the one observed in the detonation experiments on unconfined H2/air mixtures [12]. As we will show below, this mechanism proves to be substantially more efficient in the case of the considered cryogenic aerosols. Heated combustion products, mainly water, will radiate in the infrared with the maximum intensity at wavelength $\lambda_{rad} \approx (1 \div 3)\;\mu m.$ According to the infrared absorption data in [20], the peak in the absorption coefficient for LH2 occurs at wavelength $\lambda_{absorp,LH2} \approx 2.2\;\mu m$, which lies within this spectral range of radiation. At that wavelength the absorption length (the inverse of the absorption coefficient) in LH2 is $l_{absorp,LH2} \approx 3\,mm.$ Therefore, in this spectral range of the infrared the droplets are opaque and are able to absorb a significant portion of the incoming radiation. Furthermore, since $l_{absorp,LH2}$ is comparable to the droplet radius, the heat from radiation will be absorbed in the droplet bulk, raising the temperature of the liquid phase inside without significant evaporation at the droplet interface. Thus, the radiation heat may quickly raise the LH2 and LOx droplet temperature to the critical temperature $T_c$, resulting in an explosive vaporization of the entire droplet. This will greatly enhance the aerosol combustion.

The evaporation time for a droplet subject to radiative heat transfer can be estimated via the following heat balance,

$$C_L \rho_L (T_c - T_L) \left( \frac{4\pi}{3} r_{dr}^3 \right) = \eta^* \sigma T_{flame}^4 4\pi r_{dr}^2 t_{evap}, \tag{12}$$

where σ=5.67x10$^{-8}$W/m$^2$/K$^4$ is the Stephan-Boltzmann constant and $\eta^*$ is the absorption efficiency. We will estimate $\eta^* = 0.1 \div 0.5$, assuming that a significant portion of the radiation is absorbed by the droplet. We note that a more precise estimation of the absorption efficiency requires a detailed analysis of a very complex problem of light emission by the combustion products (water) and propagation through a highly heterogeneous aerosol mixture, which is beyond the scope of this paper. Here we limit ourselves to semi-quantitative estimates aimed at capturing the orders of magnitude of the quantities of interest and allowing us to fit the experimental data. As follows from Eq. (12)

$$t_{evap} = \tau_{evap} \left( \frac{T_1}{T_{flame}} \right)^4, \quad \tau_{evap} = \frac{C_L \rho_L (T_c - T_L) r_{dr}}{3\eta^* \sigma T_0^4} \tag{13}$$

where $T_1$=3500K is the characteristic flame temperature. The evaporation time strongly depends on the flame temperature $T_{flame}$. Taking $\eta^* \geq 0.25$ and the droplet radii from Section E, we have $t_{evap} \leq$ 10msec both for the LH2 and the LOx droplets at $T$ =3500K. Of course, $t_{evap}$ depends on a number of factors and may vary in the range from a few milliseconds to tens of milliseconds.

Importantly, the vaporization of the LH2 and LOx droplets results in an abrupt increase of the combustible gas density and leads to a strong buildup of pressure in the combustion products. The total mass of the burned GH2, $m_{H2}^{burned} = m_{GH2} + m_{H2,drop}$ , is controlled by the available mass of GOx, $m_{Ox}^{burned} = m_{GOx} + m_{Ox,drop} = 8m_{H2}^{burned}$ formed by the initial gaseous Ox and the evaporated LOx droplets, mixed with GH2. The product mass (water) is $m_{H2O} = m_{Ox}^{burned} + m_{H2}^{burned} = 9m_{H2}^{burned}$ and the pressure of the aerosol combustion wave can be estimated as

$$p_{acw} = \left[ 9R_{H_2O} \left( \rho_{GH2} + \rho_{H2}^{droplet} \right) + \sum_i R_i \rho_i \right] T_{flame}, \tag{14}$$

where $\rho_i, R_i$ are the mass density and the gas constants for the *i*-th unburned gas component, $\rho_{H2}^{droplet} = m_{H2}^{droplet} / V_L$, where $V_L = 2\pi L^3 / 3$ is the volume of the hemisphere in which the aerosol cloud is mixed, $p_{acw}$ is the pressure in the aerosol combustion wave. Assuming $p_{acw}$ is equal to the maximum pressure $p_{max}$ that is determined by the sensor data (Fig. 5 and Fig. 7) we can estimate the density of the droplets from Eq. (14) as

$$\rho_{O2}^{droplet} = 8\rho_{H2}^{droplet} = 8\left[\frac{p_{\max}}{9R_{H_2O}T_{flame}} - \rho_{GH2} - \frac{\sum_i R_i\rho_i}{9R_{H_2O}}\right] \approx \frac{8p_{\max}}{9R_{H_2O}T_{flame}} \quad for\ p_{\max} >> p_{atm} = 1atm \quad (15)$$

To simulate the aerosol combustion we use Eqs. (2)-(5) in which we replace Eqs. (2) with Eqs. (16):

$$\begin{aligned} &\frac{\partial C_{H2}}{\partial t} + \nabla\cdot\left(uC_{H2}\right) = -2C_{H2}C_{Ox}A(T) + \frac{C_{H2}^{droplet}}{\tau_{evap}}\left(\frac{T}{T_1}\right)^4, \\ &\frac{\partial C_{Ox}}{\partial t} + \nabla\cdot\left(uC_{Ox}\right) = -C_{H2}C_{Ox}A(T) + \frac{C_{Ox}^{droplet}}{\tau_{evap}}\left(\frac{T}{T_1}\right)^4, \\ &\frac{dC_{H2}^{droplet}}{dt} = -\frac{C_{H2}^{droplet}}{\tau_{evap}}\left(\frac{T}{T_1}\right)^4, \frac{dC_{Ox}^{droplet}}{dt} = -\frac{C_{Ox}^{droplet}}{\tau_{evap}}\left(\frac{T}{T_1}\right)^4 \end{aligned} \quad (16)$$

where $C_i$ is the molar concentration of the *i*-th component in the mixture.

As we noted above, the first HOVI group (HOVI 13 and 14) is characterized by the formation of both Ox and H2 aerosol clouds forming due to the rupture of the bottom domes in both tanks. These clouds first mix and then the explosion occurs. To simulate this aerosol explosion, we take into account that $C_{H2}^{droplet} > 0$ and $C_{Ox}^{droplet} > 0$ and have the initial values giving the pressure in the aerosol combustion wave which coincides with $p_{max}$ in the explosion wave measured by the sensors (Fig. 5 and Fig. 7). For example, according to Eq. (15) the sensor reading of $p_{max}$≈4atm in HOVI 13 corresponds to an aerosol combustion wave with $C_{H2}^{droplet} = 20 mol / m^3$ $(\rho_{H2}^{droplet} = 0.04 kg / m^3)$ and $C_{Ox}^{droplet} \approx 10 mol / m^3$ $(\rho_{ox}^{droplet} \approx 0.32 kg / m^3)$ (see Fig. 17).

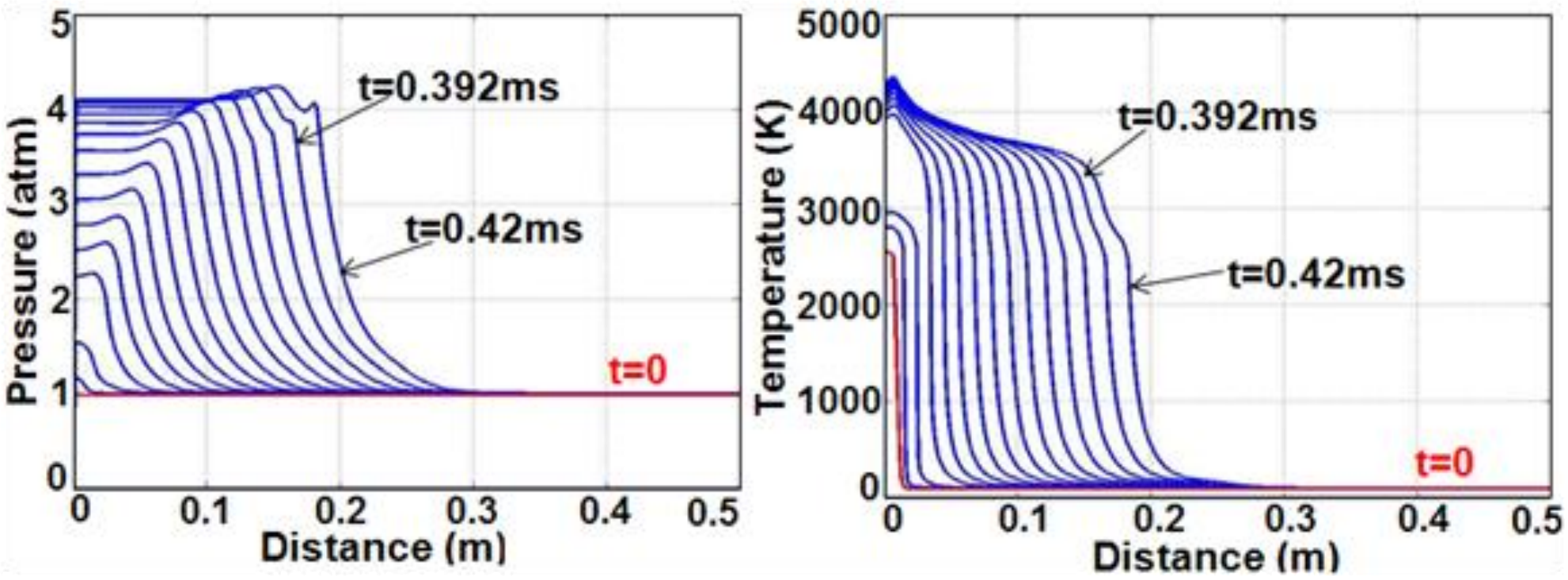

**Fig. 17 The distributions of pressure and temperature during aerosol combustion in the mixture (2:1:4) with the evaporation time $\tau_{evap} = 1ms$ and the droplet densities $\rho_{H2}$=0.04 kg/m$^3$ and $\rho_{Ox}$=0.32kg/m$^3$. Initial conditions: $p_0$=1atm and $T_0$=100K for $r> r_0$=1cm and $p_0$=1atm and $T_0$=2500K for $r < r_0$=1cm (in red). The wave parameters: $p_{max}\cong$4.2atm, $T_{max}$=3300K, $v$=600m/s.**

The second HOVI group (HOVI 2 and 5) is characterized by the formation of an Ox aerosol cloud and a gaseous GH2 cloud arising from the rupture of the LOx tank bottom dome and of the LH2 tank top dome. In this case $C_{H2}^{droplet} = 0$ and the initial concentration $C_{H2}$ in the mixture greatly exceeds $C_{Ox}$. Figure 18 shows the formation of the aerosol combustion wave with the parameters close to those of the HOVI 5 explosion (Fig. 7).

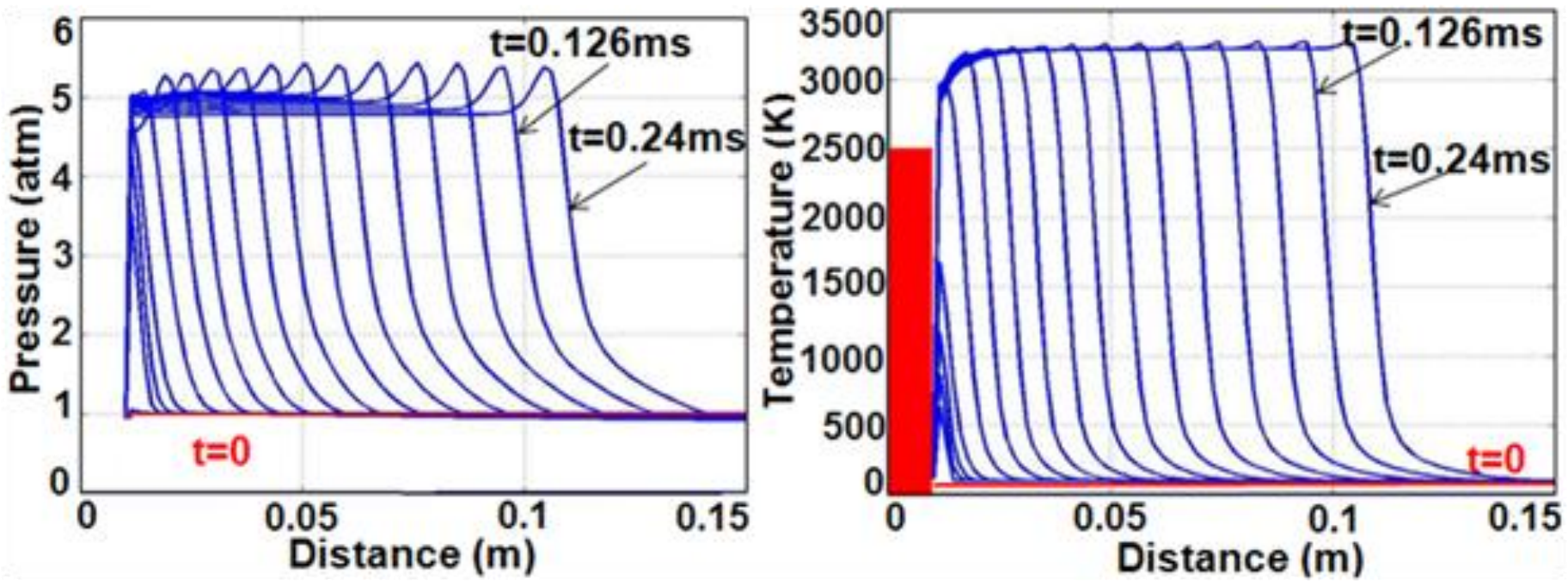

**Fig. 18 The distributions of pressure and temperature during aerosol combustion in the H2/O2/N2 (3/1/2) mixture with $\tau_{evap} = 1ms$, $\rho_{H2}$ = 0 and $\rho_{Ox}$ = 0.02kg/m$^3$. Initial conditions: $p_0$=1atm and $T_0$=100K for $r> r_0$=1cm and a hot object with $T$=2500K (in red). The wave parameters: $p_{max}\cong$ *5.4atm, $T_{max}$= 3200K, $v$= 720m/s.***

Comparison of the parameters of the simulated aerosol combustion waves with the sensor data are given in Table 1. This table summarizes our findings regarding the cryogen masses involved in different representative HOVI tests, as well as the main characteristics of the blast. Specifically, column 1 lists the numbers of the typical HOVI tests for which the analysis was presented, column 2 shows the masses of mixed LH2 and LOx aerosols found from the numerical simulations of Section F to produce a blast wave with the peak pressure observed experimentally, column 3 lists the masses of LH2/GH2 and LOx that escaped from the tanks prior to explosion computed in Section G, column 4 lists the peak pressures recorder by the sensors, column 5 lists the blast wave velocity measured experimentally, column 6 lists the values of the calculated blast wave velocity based on the shock wave speed at pressure shown in column 4, column 7 lists the duration of the dynamic pressure impulse measured by the sensors, and column 8 lists the calculated values of the blast duration computed as the ratio of the aerosol cloud radius to the shock wave velocity.

**Table 1 Main parameters of aerosol combustion and the sensor data**

| # HOVI test | Aerosol mass of mixed H2 and Ox (kg), $\tau_{evapor} = 1ms$ | Escaped H2 and Ox mass (kg) at $R_{hole} \geq 10$cm | Maximum pressure from sensor data (atm) | Experimental blast wave velocity (m/sec) $v = \Delta L / \Delta_t$ | Calculated blast wave velocity (m/sec) $v = c_0 \left(1 + \frac{(\gamma+1)\bar{p}}{2\gamma p_0}\right)^{1/2}$ | Measured duration of the blast wave (msec) | Calculated duration of the blast wave $\tau = L/v$ |
|---|---|---|---|---|---|---|---|
| 13 | ≈ 0.52 H2 ≈ 4.1 Ox | ≥ 66 LH2 ≥13 LOx | 3.3÷4.2 | ≈ 740÷825 | ≈ 760 | ≈3.5 | ≈3.5 |
| 2 | ≈ 0 H2 3.8 Ox | ≥ 0.5 GH2 ≥ 7.2 LOx | 3.2÷5.4 | ≈ 780 | ≈ 785 | ≈3.4 | ≈3.4 |
| 5 | ≈ 0 H2 2.7 Ox | ≥ 0.9 GH2 ≥ 7.1 LOx | 2.5÷3 | ≈ 660 | ≈ 660 | ≈4.0 | ≈4.0 |
| 9 | 8.3÷15.2 H2 66-121 Ox | ≥6.2 LH2 ≥ 23 LOx | 80÷110 | 2625÷2966 | 2500÷2928 | ≈0.7 | ≈0.7 |

One can see from these data that only a small fraction of the escaped cryogens mass participates in the explosion in the form of aerosol. Also, the blast wave velocity and duration agree very well with the numbers based on the peak pressure and the size of the aerosol cloud. We note that deflagration wave acceleration in aerosol H2/O2/N2 mixtures also occurs for lower droplet evaporation rates, i.e., for $\tau_{evap} > 1ms$. However, for larger values of $\tau_{evap}$ one obtains very similar combustion waves, provided that initial droplet mass is higher. For example, for the 2:1:4 mixture and $\tau_{evap} = 5m\sec$ one finds $\rho_{H2}$=0.07kg/m$^3$ and $\rho_{Ox}$=0.56kg/m$^3$. Let us point out that an aerosol deflagration wave can accelerate and transform into a strong detonation wave for large enough aerosol densities and sizes of the mixed H2 and O2 aerosol clouds. This effect is considered in Section G.

**G. Estimates of the escaped H2 and Ox masses before the explosion**

The aerosol combustion is determined by the H2 and Ox masses inside the overlapping area of the H2 and Ox aerosol clouds. These masses are $m_{H2}^{aerosol} = \rho_{H2}^{aerosol}\left(2\pi/3\right)L_{over}^3$, $m_{Ox}^{aerosol} = \rho_{Ox}^{aerosol}\left(2\pi/3\right)L_{over}^3$. The typical size of the overlap area is $L_{over}\approx 2$m. Obviously the H2 and Ox masses escaped from the ruptured tanks during the short delay time $t_{delay}<100$msec before the explosion have to be greater than the aerosol masses obtained in Section F. Our estimates below show that this condition is fulfilled for all HOVI tests. Indeed, the escaped LH2 mass in the first group of the HOVI tests can be estimated from the equations of momentum, energy and mass conservation for the LH2 flow:

$$p_{at} - \frac{\rho_L v_0^2}{2} = p - \frac{\rho_L v^2}{2}, \ \frac{p}{p_0} = \left(\frac{V_{g0}}{V_g}\right)^{\gamma}, \quad \gamma = \frac{C_p}{C_v} = 1.4$$
$$dh_g = -dh_L = \frac{S_h}{S_0} v dt, \ dV_g = -S_0 dh_g = S_h v dt = -dV_L \tag{17}$$

where $S_h = \pi R_h^2$ is the cross-section area of the hole in the tank bottom. It follows form Eq. (17) that

$$\frac{dV_g}{dt} = a\left(\left(\frac{V_{g0}}{V_g}\right)^{\gamma} - \frac{p_{atm}}{p_0} + \frac{\rho_L v_0^2}{2p_0}\right)^{1/2}, \ a = S_h\left(\frac{2p_0}{\rho_L}\right)^{1/2} \tag{18}$$

For LH2 we have $\rho_L = 70kg/m^3, \gamma = C_p/C_v = 1.4, p_0 = 1.45atm, p_{atm} = 1atm$ and the condition $\frac{\rho_L v_0^2}{2p_0} < \frac{p_{atm}}{p_0}$ is valid for the impact velocity $v_0 <$ 45m/sec. Therefore, according to Eq. (18)

$$V_{L,\max} = V_{g0}\left(\frac{p_{atm}}{p_0} - \frac{\rho_L v_0^2}{2p_0}\right)^{-1/\gamma}. \tag{19}$$

It then follows that $m_{LH2}^{escaped} = \rho_{LH2} V_{g,\max} \simeq 66kg$ for HOVI 13 ($V_{g0}=0.553m^3$) and $m_{LH2}^{escaped} \simeq 21kg$ for HOVI 14 ($V_{g0}=0.176m^3$). The volume of the escaped LH2 equals to its maximum value $V_{L,\,max}$ for time $t\approx$20ms < $t_{delay}$. $V_{L,\,max}$ does not depend on the ignition delay time $t_{delay}$ and the cross-section $S_h = \pi r_h^2$ of the hole in the LH2 tank top. We see that the values of $m_{LH2}^{escaped}$ for HOVI 13 and 14 are much more than the necessary mass of the H2 aerosol in the overlapping H2 and Ox clouds: $m_{H2}^{droplet} = \rho_{H2}^{droplet} V_{H2}^{droplet} = 0.67kg,\ \ V_{H2}^{droplet} = (2\pi/3) L_{overlap}^3$ for $\rho_{H2}^{droplet} = 0.04kg/m^3$ *and* $L_{overlap} = 2m.$ For Ox, we have $\rho_L = 1141kg/m^3, \gamma = C_p/C_v = 1.4,\ p_0 = 2.1atm,$ $p_{atm} = 1atm$ and the opposite condition $\frac{\rho_L v_0^2}{2p_0} > \frac{p_{atm}}{p_0}$ is valid for the impact velocity $v_0$>14m/s. Analysis of Eq. (18) shows that $m_{LOx}^{escaped} > 15kg$ for the ignition delay time $t_{delay}$> 50msec and $R_h$> 10cm. At the same time the droplet mass $m_{Ox}^{droplet} = \rho_{Ox}^{droplet} V_{H2}^{droplet} = 5.4kg < m_{Ox}^{escaped}$. Moreover, there is additional oxygen in the atmosphere.

The second group of the HOVI tests is characterized by the rupture device located between the tanks. In this case the impact results in the rupture of the bottom of the LOx tank and the top of the LH2 tank. As a consequence, liquid Ox and gaseous H2 escape from the tanks. The escaped LOx masses are determined by Eq. (18) and their values are presented in Table 1. The escape dynamics of GH2 is given by [21]

$$V_{g0}\frac{d\rho_v}{dt} = -jS_h,\ \ p = R_v \rho_v T,$$
$$j(t) = \left(\frac{p_0}{p}\right)^{1/\gamma} \sqrt{\frac{2\gamma}{\gamma-1} p\rho_v \left(1 - \left(\frac{p_0}{p}\right)^{(1-\gamma)/\gamma}\right)} \tag{20}$$
$$m_{H2} = S_h \int_0^{t_{esc}} j(t)\, dt,\ \ m_{H2}^{\max} \approx \frac{p(0) - p_0}{p_0} \rho_v V_{g0}$$

The values of the escaped GH2 mass for the delay time $t_{delay} \approx$ 100 msec are presented in Table 1. These parameters were taken into account in other calculations.

### H. Deflagration-to-detonation transition in aerosol mixtures

As we noted above, deflagration in the explosive H2/Ox mixtures may be initiated by a small spark or any hot object with temperature $T > T_{ign}$ (Fig. 14). Due to the presence of aerosols the deflagration wave accelerates and its pressure increases (Section F). When the pressure exceeds the value ~$p_{cj}$, a strong detonation wave arises. The characteristics of this wave are determined only by the parameters of the H2/Ox mixture and do not depend on the initial condition triggering this explosion. Thus, the greatest explosion risk is posed by the aerosol H2/Ox mixtures. These mixtures are easily ignited and can produce an especially strong explosion when the mixtures are close to the stoichiometric composition. This effect is realized at relatively high droplet densities of both H2 and Ox components (Fig. 19), or-even LOx alone (Fig. 20).

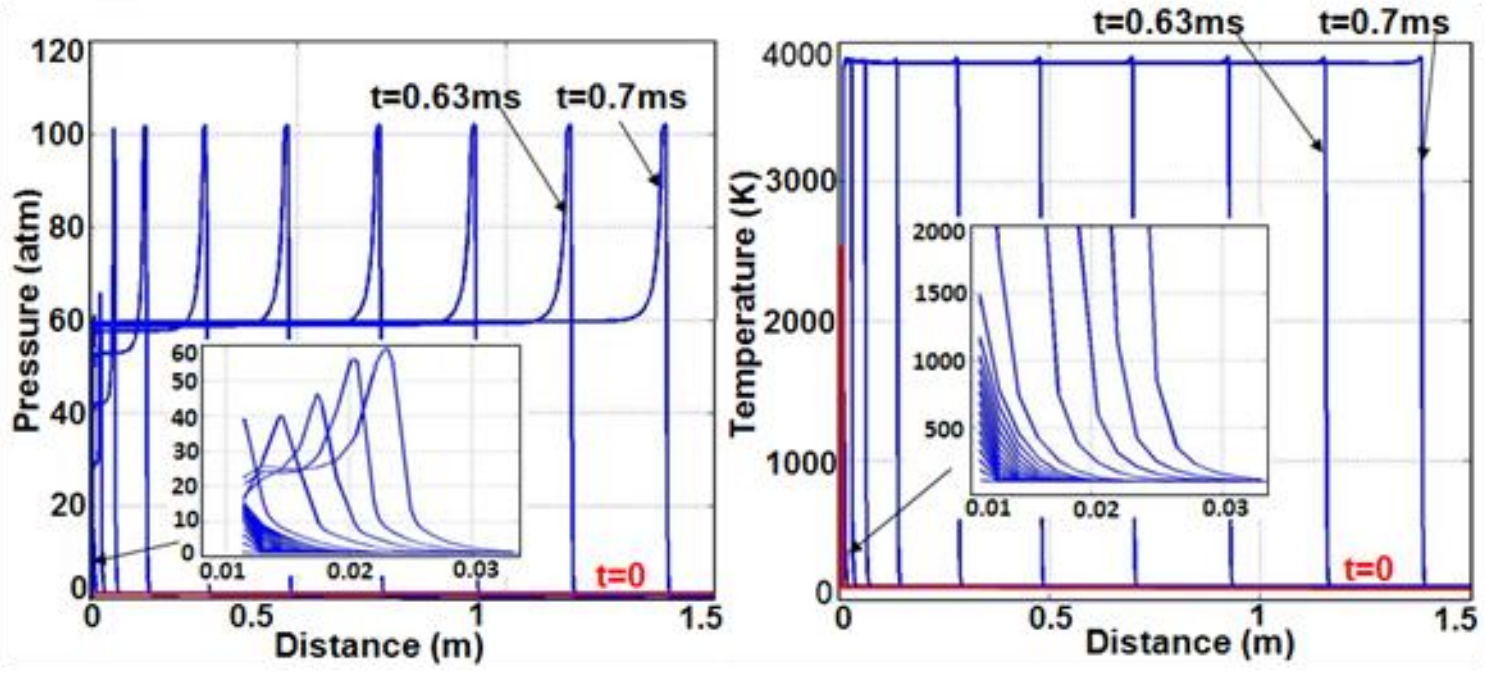

**ig. 19 Deflagration-to-detonation transition in the aerosol H2/O2 mixture (~2:1) with the evaporation time $\tau_{evap} = 1ms$ and the droplet densities $\rho_{H2}$=0.08 and $\rho_{Ox}$=0.64kg/m3. Initial conditions: $p_0$=1atm and $T_0$=100K for *r*> *r0*=1cm and *T*=2500K (in red). The wave parameters: $p_{max} \cong$ *102atm*, $T_{max}$= *3800K, v =3000m/s.* Inserts show early stages of the processes.**

This leads us to an important conclusion: a strong detonation explosion can be initiated by a spark or a hot object in an aerosol mixture, not just by a strong shock wave, as is the case for cryogenic gaseous mixtures (Section B). We note that the maximum pressure of the aerosol detonation wave can exceed that of the detonation wave in gas mixtures (compare Fig. 11 with Fig. 19). The aerosol detonation scenario is apparently realized in the HOVI 9 test, which resulted in an unusually strong blast (see detail in Section E).

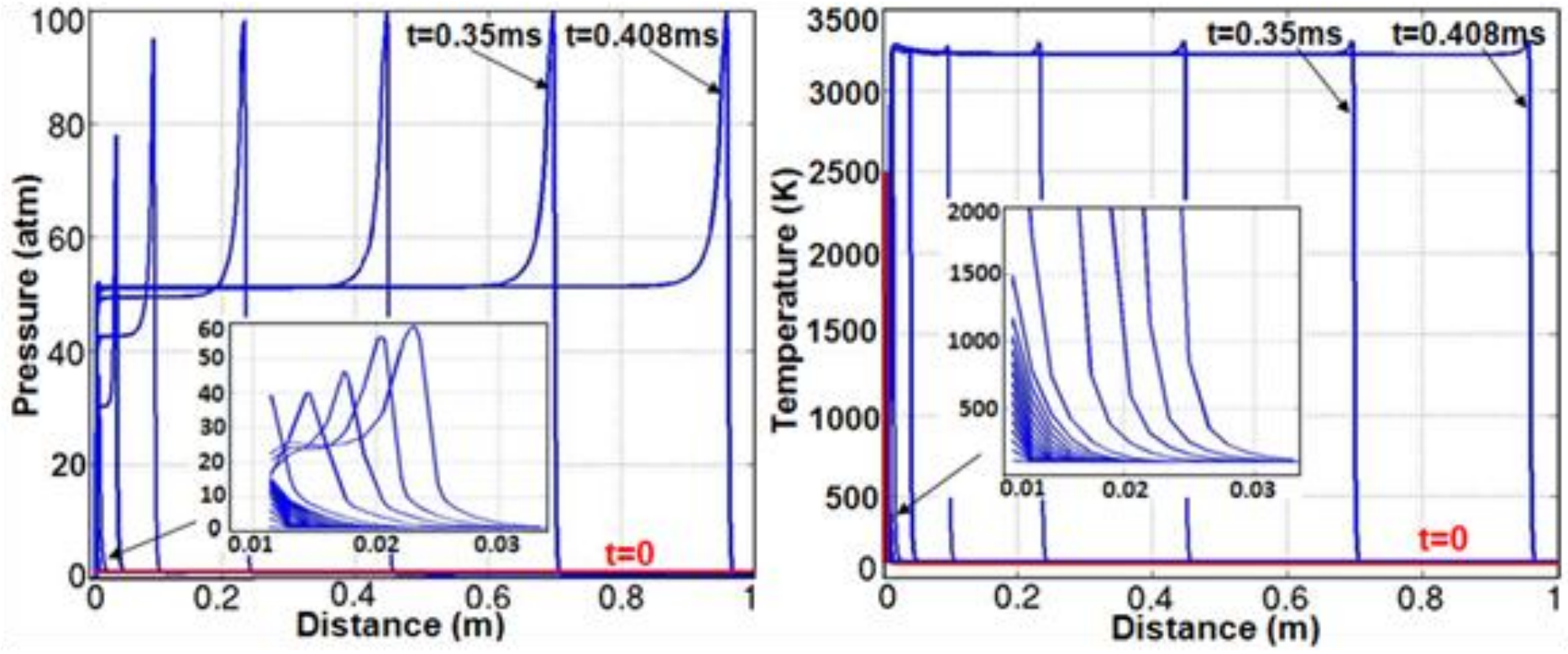

**Fig. 20 Deflagration-to-detonation transition (3:1:0) in the aerosol H2/O2/N2 mixture with $\tau_{evap} = 1ms$, $\rho_{H2}$=0 and $\rho_{Ox}$=0.48kg/m3. Initial conditions: $p_0$=1atm and $T_0$=100K for r> r0=1cm and T=2500K (in red). The wave parameters: $p_{max} \cong$ *100atm*, $T_{max}$ *=3800K, v =3000m/s.* Inserts show the beginning stage of the processes.**

Let us point out that the effect of deflagration-to-detonation transition in aerosol H2/O2/N2 mixtures discussed above also occurs at lower rates of droplet evaporation, e.g., when the evaporation time $\tau_{evap} > 1ms$. However, for larger values of $\tau_{evap}$ higher droplet mass is necessary. For example, for stoichiometric aerosol H2/O2 mixtures one needs $\rho_{H2}$=0.32kg/m$^3$ at $\tau_{evap} = 5ms$.

## IV. Cavitation-induced ignition of H2/O2/N2 mixtures

One of the puzzling questions in the studies of cryogenic explosions is the mechanism of ignition in the cryogenic H2/O2/N2 mixtures [22]. Multiple tests, including the HOVI tests, give clear evidence that the ignition in these mixtures is not due to external sources. GH2 released alone does not explode, despite of the large amount of oxygen present in the air. Ignition occurs when gaseous or aerosol cryogenic hydrogen and oxygen, as well as liquid oxygen (LOx) are available. The tests showed that self-ignition of the cryogenic mixtures is realized always for the short time less than 0.1s after GH2 and GOx flows are mixed with a turbulent LOx stream. This condition is fulfilled in all HOVI tests. Below we discuss a possible mechanism of self-ignition of cryogenic H2/Ox mixtures that can be realized when GH2, GOx, and LOx flows mix.

### I. Cavitation-induced Detonation of H2/O2/N2 Mixtures

We propose a cavitation-induced mechanism which may explain self-ignition in cryogenic H2/Ox mixtures. Cavitation refers to the formation and compression of vapor bubbles in a liquid under the action of the applied pressure difference. Due to the inertial motion of the liquid, this process results in a rapidly collapsing bubble and an increase in the gas temperature and pressure inside the bubble, producing a strong shock wave [23, 24]. In the simplest case, the dynamics of the bubble radius is described by the Rayleigh-Plesset equation [22]

$$R_b \frac{d^2 R_b}{dt^2} + \frac{3}{2}\left(\frac{dR_b}{dt}\right)^2 + \frac{4\mu_L}{\rho_L R_b}\frac{dR_b}{dt} + \frac{2\sigma}{\rho_L R} = \frac{p_b(R_b) - p_{L\infty}}{\rho_L} \qquad (21)$$

where $p_b(R_b)$ and $p_{L\infty}$ are the pressures inside the bubble and in the liquid far from it, respectively. A pressure jump $p_b(R_b) - p_s$ between the LOx and the bubble can arise from the turbulent mix of gaseous H2 and Ox and Ox liquid stream or due to relatively weak initiating shock wave arising from in the LOx piece containing the bubble as a result of impact of this piece on a solid object. Cavitation-induced ignition of H2/Ox mixtures may be caused by the second strong shock wave generated by a collapsing bubble near the LOx surface which propagates across the GH2/GOx-LOx interface and ignites the gas mixture (Fig. 21).

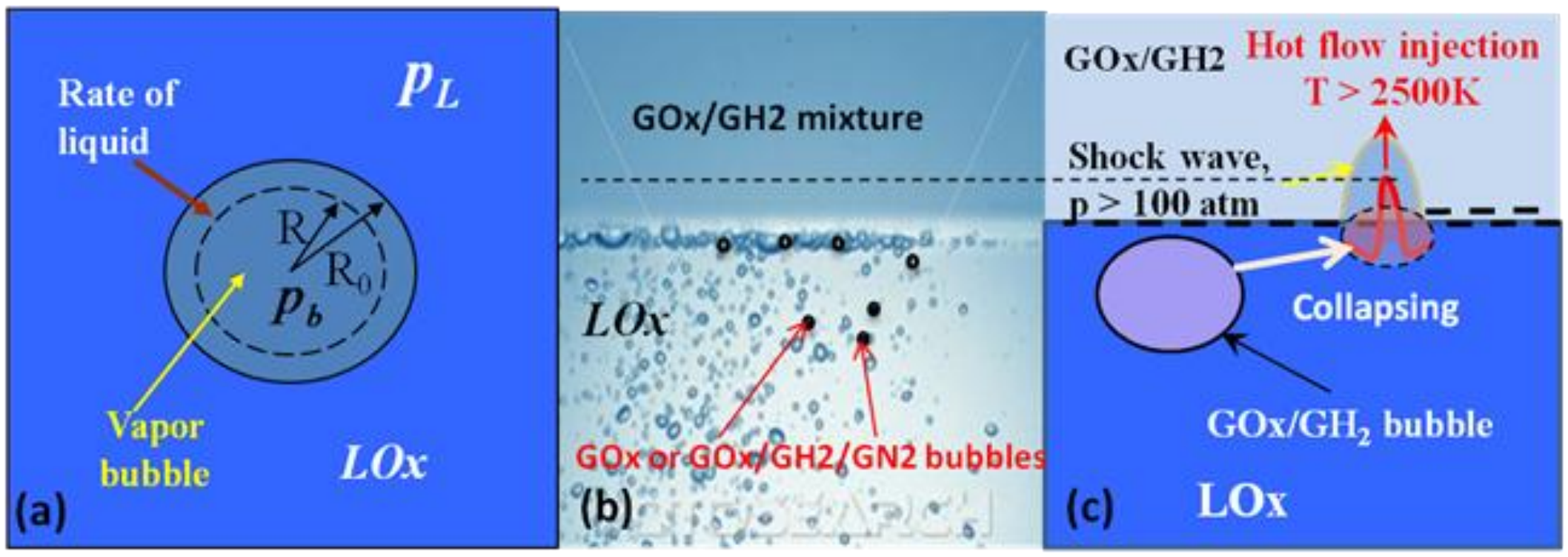

**Fig. 21 Cavitation-induced ignition: collapse of a vapor Ox bubble in liquid Ox under the action of a pressure differential $p_L$-$p_b$ (a); the bubble can contain only Ox vapor or a GOx/GH2/GN2 mixture (b); generation of the shock wave and injection of hot gas from a bubble collapsing near the LOx - GOx/GH2 interface (c).**

To estimate the gas parameters in the collapsing bubble, we will assume that the bubble contains Ox vapor and maybe a small portion of GH2. The analysis carried out in [23, 24] show that we can assume that the Ox vapor pressure due to active condensation is closed to the pressure of saturated Ox vapor, and the compression of GH2 in the collapsing bubble is an adiabatic process. In other words, we can write the pressure in the bubble as [23]

$$p_b(t) = p_S(T_S) + p_{g0}\left(R_{b0} / R\right)^{3\gamma} \tag{22}$$

Here $R_{b0}$ and $p_{g0}$ are the initial bubble radius and the initial partial pressure of the H2 gas in the bubble, $p_S(T_S)$ is the saturation pressure of the Ox vapor at temperature $T$ equal to the temperature $T_s$ at the liquid-vapor interface. The time of the bubble collapse can be estimated by [22]

$$t_{TC} = 0.915\left(\frac{\rho_L R_{b0}^2}{p_{L\infty} - p_s}\right)^{1/2} \tag{23}$$

The results of the numerical analysis of Eqs. (21)-(22) are presented in Fig. 22. These results show that the collapse time is close to that given by Eq. (23). We also see that under the action of the pressure difference exceeding 0.2 atm the initial bubble radius $R_0$=2mm goes to its minimum value $R_{min}$≈0.1mm, and the temperature and pressure in the collapsing bubble can exceed 500 atm and 2500K, respectively. Our simulations based on Eqs. (2)-(5) show that a localized shock wave of lower intensity is sufficient to induce detonation both in stoichiometric H2/Ox (Fig. 23) and

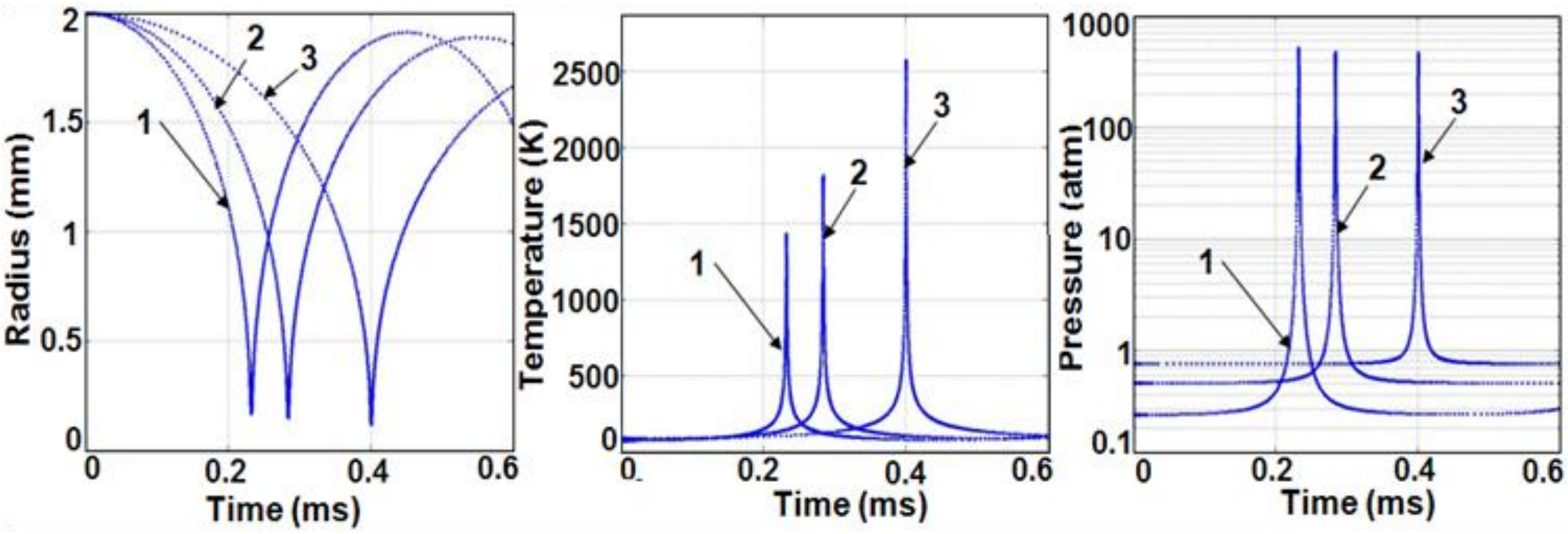

**Fig. 22 Collapse of a GOx/GH2 bubble of initial radius $R_{b0}$=2mm in LOx. Initial conditions: 1 - $p_L$-$p_{O2}$=0.75atm, $p_{O2}$=0.25atm, $p_{H2}$=0.015atm; 2 - $p_L$-$p_{O2}$=0.5atm, $p_{O2}$=0.5atm, $p_{H2}$=0.0085atm; 3 - $p_L$-$p_{O2}$=0.25atm, $p_{O2}$=0.75atm, $p_{H2}$=0.003atm. The parameters are: 1 - $p_{max}$=510atm, $T_{max}$=1450K, $R_{bmin}$=0.165mm, collapse time $t_{collapse}$=0.23ms: 2 - $p_{max}$ =460atm, $T_{max}$=1850K, $R_{bmin}$=0.15mm, $t_{collapse}$= 0.284ms; 3 - $p_{max}$ = 450atm, $R_{bmin}$= 0.11mm, $T_{max}$=2500K, $t_{collapse}$= 0.4ms.**

H2/Ox/N2 mixtures (Fig. 24). We note that in real situation the super-hot and super-compressed O, H, OH species are formed in process of mixture burning [6] in the bubble (see Fig. 8). These species will be ejected from the bubble into the space above the LOx surface and easily ignite the GH2/GOx mixture situated under it.

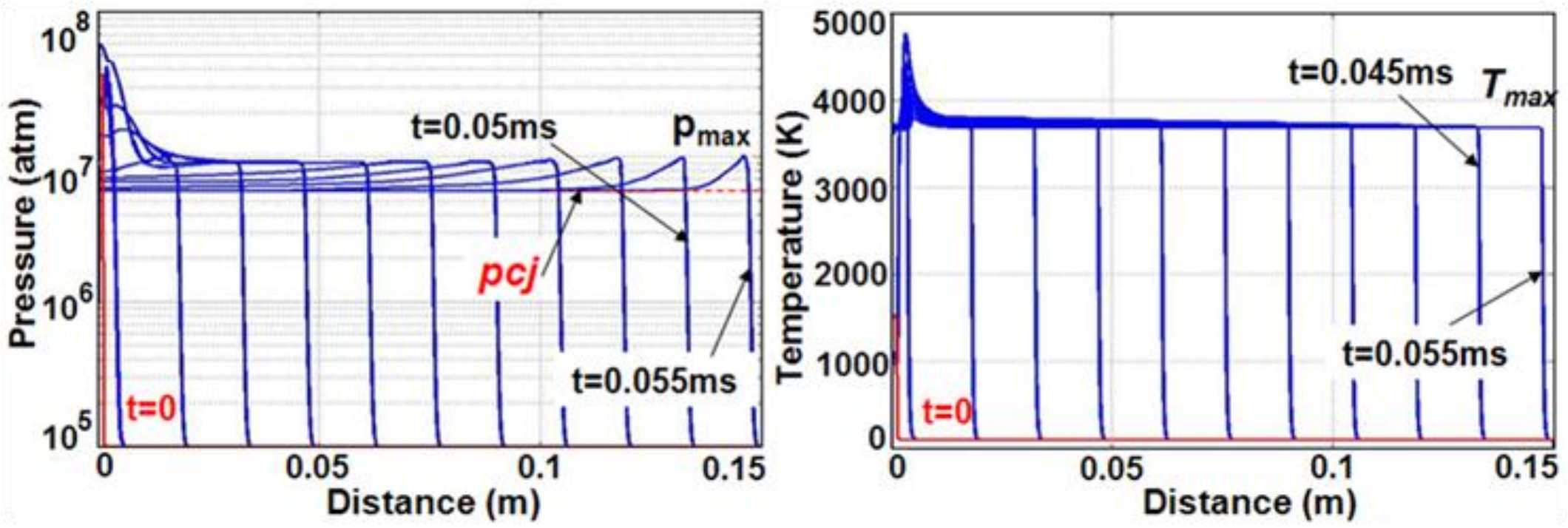

**Fig. 23 Cavitation-induced ignition of a stoichiometric H2/Ox mixture (2:1) with the mixture temperature $T$=100K and pressure $p$=1atm for $r$>0.15msec. Initial cavitation condition: temperature $T_0$=1500K and pressure $p_0$=350atm for $r$<0.15mm (in red). The detonation wave parameters: $p_{max}$=100atm, $p_{cj}$=60atm, $T_{max}$=3800K, $v_{dw}$=3000m/s.**

The main requirement for the cavitation onset is a fast jump of pressure between the liquid and the vapor bubble. This jump may be of different nature. We list here possible scenarios of cavitation-induced ignition in H2/O2/N2 mixtures:

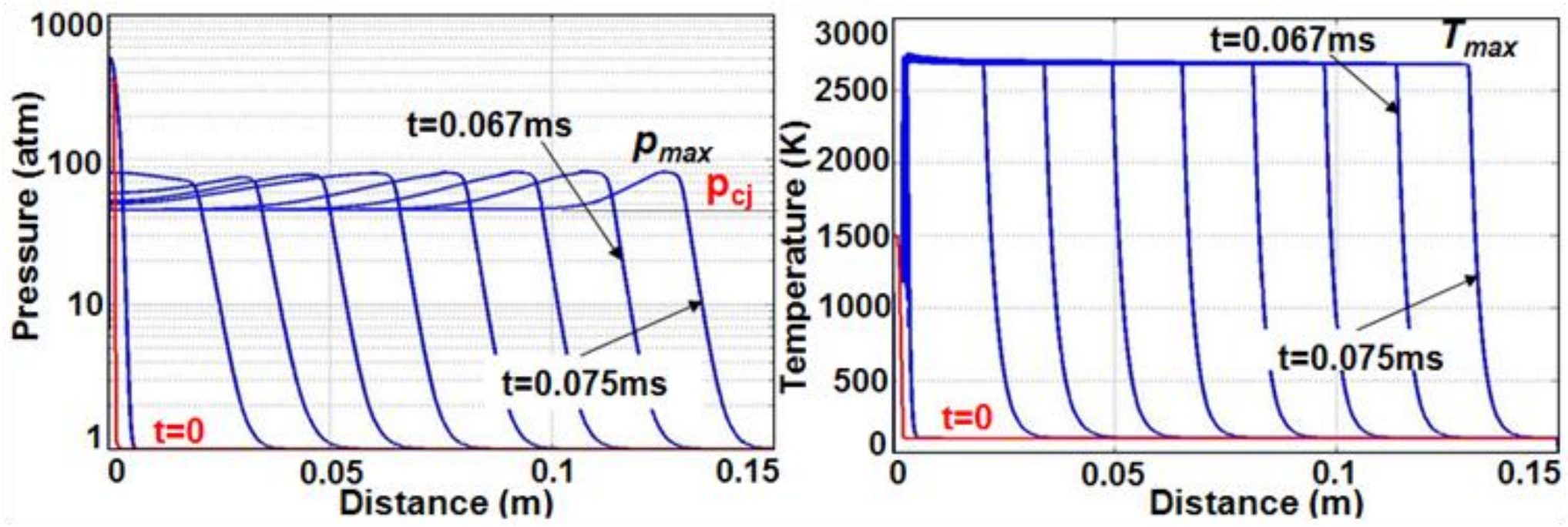

**Fig. 24 Cavitation-induced ignition of a H2/Ox/N2 mixture (2:1:4) with mixture temperature $T$=100K and pressure $p$=1atm for $r$>0.15mm. Initial cavitation condition: temperature $T_0$=1500K and pressure $p_0$=350atm for $r$<0.15mm (in red). The detonation wave parameters: $p_{max}$=82atm, $p_{cj}$=45atm, $T_{max}$=2800K, $v_{dw}$=2000m/s.**

. *1. Collapse of a rarefied vapor bubble.* Rarefied vapor bubbles can arise in the liquid as a result of turbulent mixing of GOx and LOx streams and collapse in it.

*2. Collapse of a vapor Ox bubble with admixed GH2.* In this case ignition can be intensified due to the possibility of a local explosion inside the collapsing bubble because of the high temperature and pressure in it. Along with the generation of a strong shock wave, the super-hot and super-compressed O, H, OH species can form in the bubble and be ejected from it into the space above the LOx interface and easily ignite the GH2/GOx mixture.

3. *Formation of bubbles with chilled surface.* The surfaces of large LOx pieces can be cooled by very cold GH2 flows. The Ox vapor bubbles with a thin surface layer of cooled liquid and, probably, with admixed gaseous H2 as well, can rapidly form as the result an of impact of two LOx pieces. Due to the low surface temperature the pressure in the bubble quickly drops because of intense vapor condensation inducing bubble collapse.

*4.Injection of an LH2 droplet into LOx.* The impact of the LH2 and LOx jets with the ground results in turbulence and fragmentation of the liquids into droplets. A cold LH2 droplet (with $T$ = 20K) can be captured by a “hot” LOx piece ($T$=90K). The LH2 droplet will then evaporate very quickly and the pressure inside the bubble will quickly grow and can become much greater than the pressure $p_L$ in LOx. As a result, the bubble radius will increase. Due to the inertial motion of the liquid the bubble expansion can lead to the final pressure much less than $p_L$. Afterwards, the bubble starts collapsing and the pressure and temperature of the GH2/GOx mixture inside the bubble can become very high and initiate a strong shock wave.

These scenarios were confirmed by our preliminary numerical simulations. A detailed analysis of these results is beyond the scope of this paper [25].

### J. Cavitation-induced deflagration of inhomogeneous H2/O2/N2 mixtures

H2/Ox/N2 explosive mixtures containing gaseous and/or aerosol hydrogen, oxygen, and nitrogen and LOx may have strongly inhomogeneous. The collapse of a bubble near the LOx interface can generate a strong shock wave of high pressure and temperature, inducing a detonation wave in a hydrogen-rich area located near the LOx surface. If this area is surrounded by a region with low H2 density (Fig. 25), then the detonation wave will rapidly dissipate in this region and the temperature wave can initiate a deflagration in the mixture at the other side of the hydrogen-depleted region.

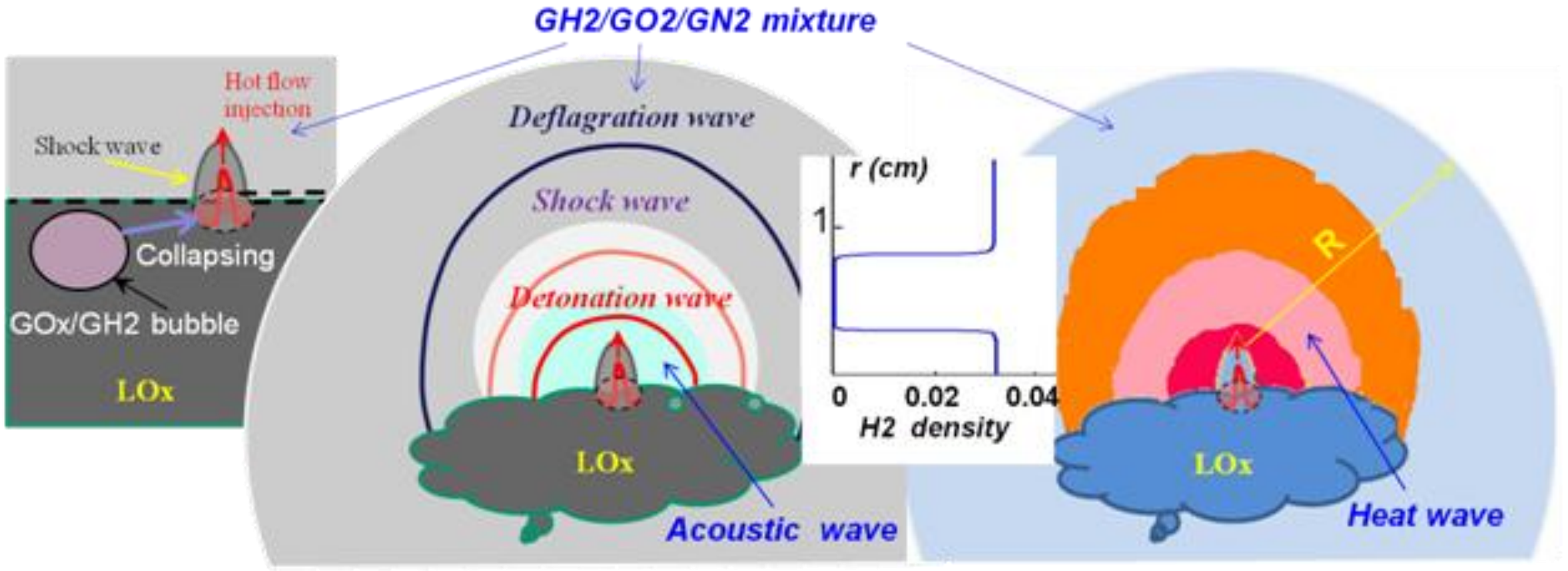

**Fig. 25 Transition from local detonation to deflagration: a sketch of the propagating blast and heat waves in a GH2/GO2/GN2 mixture generated by a collapsing bubble.**

Our simulations confirm this conclusion (Fig. 26). We see that the initial parameters of the explosion wave change from pressure $p$=350 atm to 1 atm, from temperature $T$=3000K to 2600K, and the front velocity from $v$=1200m/sec to 30m/sec in the forming deflagration wave.

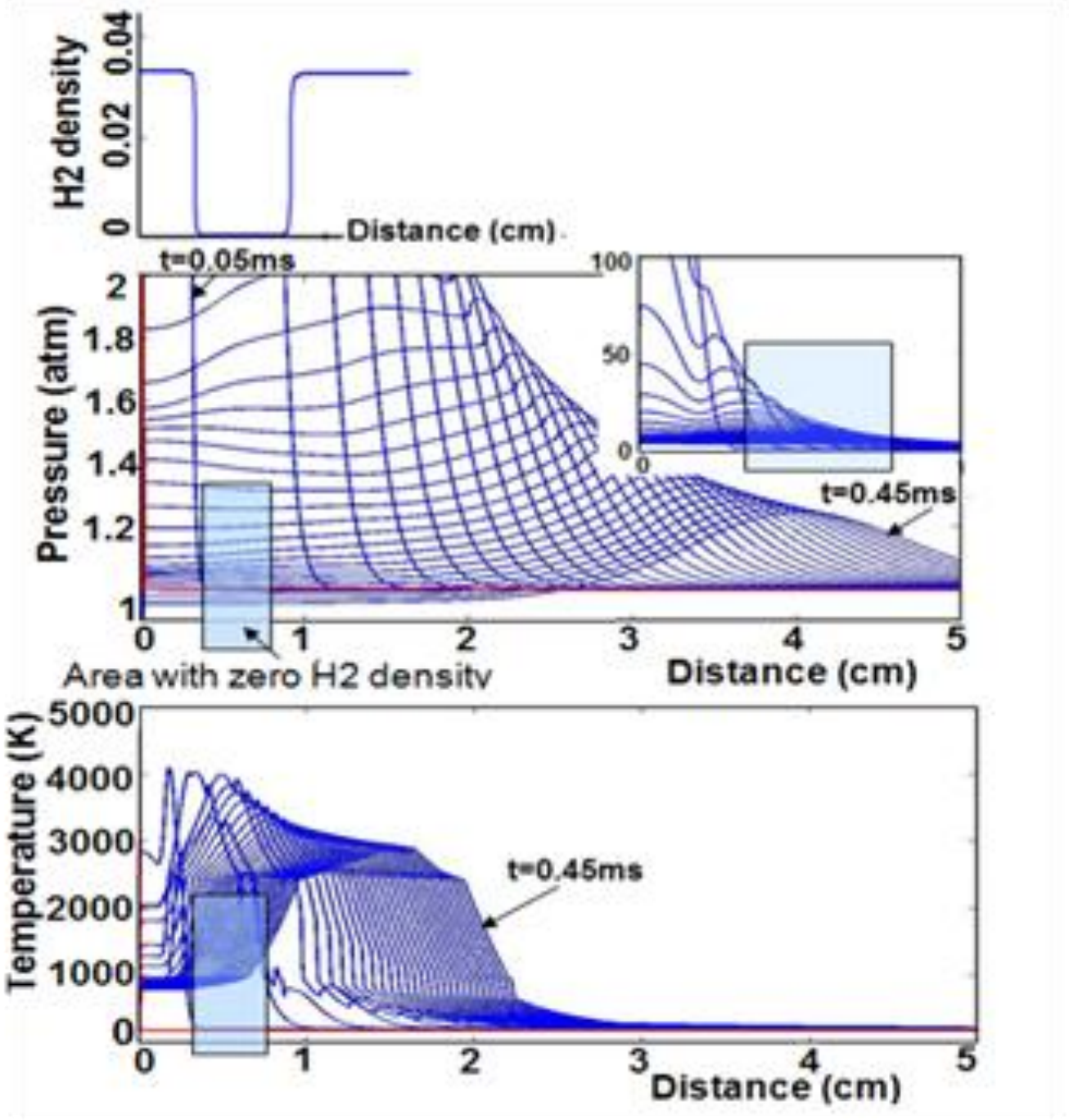

**Fig. 26 Transition from a cavitations-induced local detonation to deflagration of a GH2/GOx mixture with *T*=100K and *p*=1 atm. Initial condition: *T*=3000K, *p*=350atm inside area $r<r_0$=0.1mm. The parameters of the deflagration wave: $T_{max}$=*2600K, p≅1 atm,* v=*30m/s*.**

## V. Conclusion: conditions and risks of the onset of strong blasts in the cryogenic mixtures

We believe that proposed cavitation-induced ignition mechanism and considered scenarios of its realization of are important for understanding conditions and risk of explosion of cryogenic H2/Ox fuel used in liquid rockets and other vehicles.

The main results of our analysis are summarized in Fig. 27 describing possible scenarios and conditions of different combustion types for cryogenic H2/Ox/N2 mixtures. The combustion type is determined by the composition of the mixtures, the degree of their mixing, and also by the ignition mechanism. The *strong explosion* (detonation) can arise in gaseous GH2/GOx/GN2 mixtures when they are well-mixed and the ignition is induced by a strong shock. The closer the mixture to the stoichiometric composition, the stronger the detonation. Any other ignition source, e.g. by a spark or a hot object, results in *slow combustion* (deflagration) in gaseous GH2/GOx/GN2 mixtures.

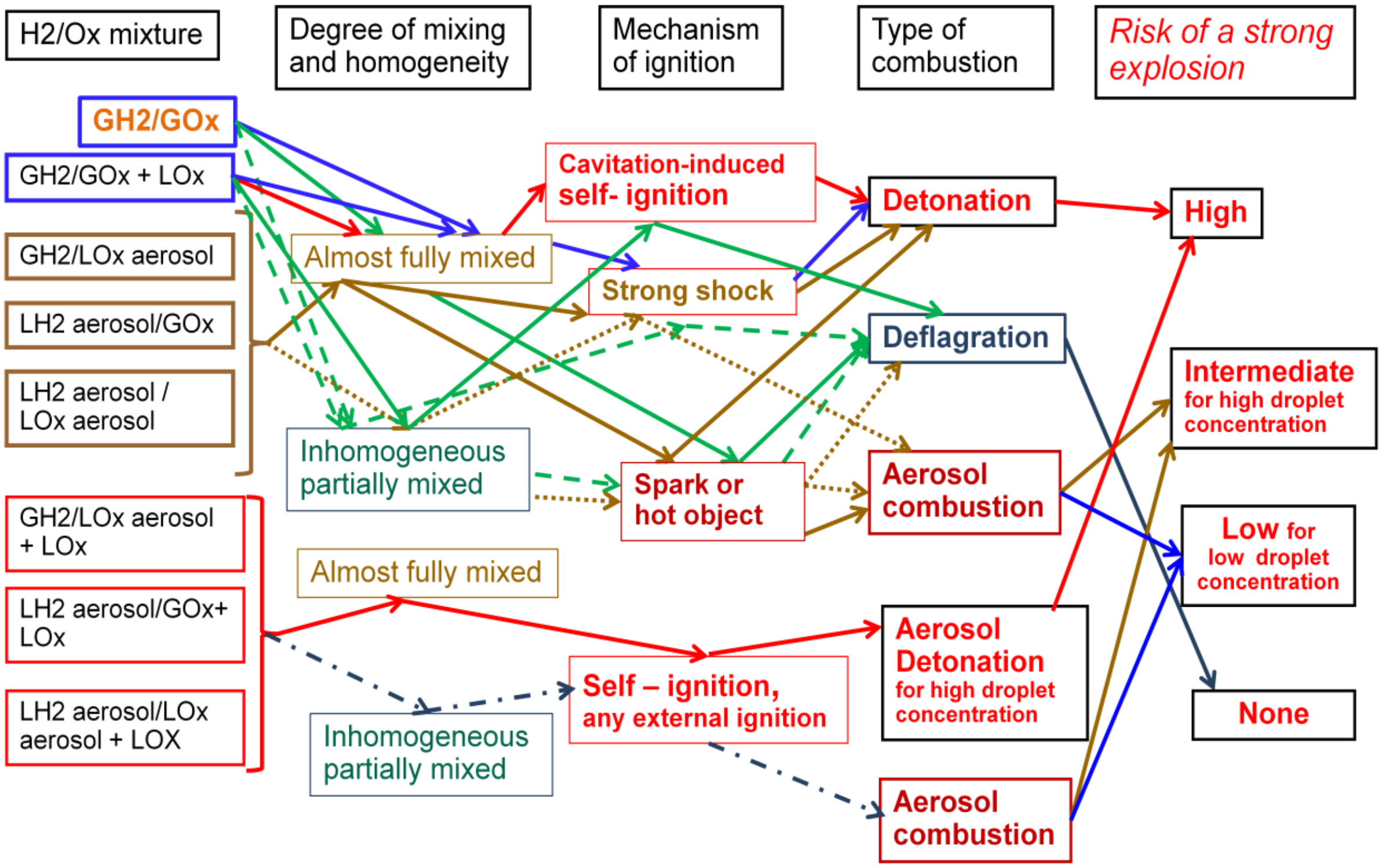

**Fig. 27 Diagram of possible cryogenic H2/Ox explosion scenarios.**

Self-ignition is realized when GH2/GOx/GN2 flows mix with a turbulent LOx stream. Any mechanism of ignition of partially mixed aerosol H2/Ox mixtures with relatively low droplet concentration leads to aerosol combustion, accelerated deflagration, and is characterized by the overpressure of several bars. *Well-mixed H2/Ox aerosol mixtures with high droplet concentrations are most dangerous*: any ignition mechanism, including self-ignition, in contact with relatively large LOx pieces results in a strong explosion. The maximum pressure in such an explosion may exceed 100 atm for stoichiometric mixtures.

The explosive power depends on the pressure and the size of the area of mixed H2 and Ox aerosol clouds. A strong detonation arises when the mixture is well-mixed inside a large enough volume. The blast intensity may be characterized by the dynamic pressure impulse $p_{max}\tau_{imp}$, where $\tau_{imp}=L/v_{dw}$ is the blast impulse duration and $L$ is the size of the mixed clouds. This quantity may be used to determine the TNT equivalent of the explosion [4]. The value of $L$ depends on the mixing time of the aerosol clouds, i.e. the time of delay $t_{delay}$ between the tank breach and the explosion. The relatively small value of $t_{delay} \approx (30\div60)msec$ in the HOVI tests corresponds to the case when the breaches occur in the intertank area and gaseous H2 and liquid Ox are injected into the intertank area from the upper ruptured LH2 tank dome and lower ruptured LOx tank dome. Such a situation was realized by accident in the Challenger disaster [1-3] and purposely in the second group of the HOVI tests (Fig. 2c). The escaped GH2 and LOx flows very quickly (but poorly) mixed together and self-ignited. Similar events can occur in the case of the bulkhead failure [26].

The aerosol combustion, accelerating the deflagration combustion, is likely to arise in the cases with the maximum blast pressure from one to several bars. Here the risk of a strong explosion is intermediate. Different situation may occur when the bottom domes of both tanks are ruptured, as can be seen from the first group of the HOVI tests (Fig. 2a and 2b). Here the delay can be longer due to a much larger area separating the H2 and Ox aerosol clouds and the large LOx pieces. Such an event was probably realized in the HOVI 9 test. This test used doubled tanks (Fig. 2b) with the rupture devices placed underneath each tank. Following the rupture of the LH2 tank, its thermal insulation detached and shielded the H2 aerosol cloud from direct contact with the turbulent LOx stream. This resulted in a long delay time $t_{delay} \approx 200\ msec$ between the rupture and the cavitation-induced ignition. During this time, the expanding H2 and Ox aerosol clouds had a chance to form and mix together. As a consequence, a nearly stoichiometric H2/Ox mixture confined to a hemi-sphere of radius of about 2m was formed. The self-ignition of this mixture induced the strongest explosion (aerosol detonation), characterized by the maximum pressure exceeding 100 atm and the duration of about 1msec (Fig. 24). We attribute this strong explosion to a chance occurrence of the complex breaking dynamics of the insulating foam detaching from the double LH2 tank. Such a strong explosion was not observed in all other 13 HOVI tests, including the HOVI 10 test, which had the same double tanks as HOVI 9. In the case of HOVI 10 the value of $t_{delay}$ was very small and $p_{max}$=1.36 atm.

An event in which the contents of the LH2 and LOx tanks is scattered over a relatively large area is probably the most dangerous from the point of view of the risk of the strongest explosion. Indeed, in this case the escaped H2 and Ox liquids will have large enough time to evaporate and to generate GH2 and GOx aerosol clouds that will have time to mix together into a large enough area before self-igniting upon contact with the ejected turbulent LOx stream. We emphasize once more that self-ignition of cryogenic H2/O2 mixtures is always realized when GH2 and GOx flows are mixed with a turbulent LOx stream.

We acknowledge Dr. Frank Benz from NASA Johnston Space Center White Sands Test Facility for providing us the test data for Hydrogen Oxygen vertical impact tests performed at the facility. We also acknowledge Dr. Benz for interesting discussions.